\documentclass[%
reprint,
superscriptaddress,
 amsmath,amssymb,
 aps,
]{revtex4-1}

\usepackage{float}
\usepackage{graphicx}% Include figure files
\usepackage{dcolumn}% Align table columns on decimal point
\usepackage{bm}% bold math
\usepackage{inputenc}
\usepackage{gensymb}
\usepackage{lmodern}
\usepackage{makecell}
\usepackage{color}
\usepackage{multirow}

\usepackage{hyperref}% add hypertext capabilities
\hypersetup{
colorlinks=true,% false: boxed links; true: colored links
linkcolor=blue,% color of internal links
citecolor=blue,% color of links to bibliography
urlcolor=blue% color of external links
}

\graphicspath{{./images/}}

\usepackage{booktabs}

\usepackage{xcolor}%Added by A.O.
\usepackage{mathrsfs}
\newcommand{\nm}{\nonumber}

\newcommand{\sslash}{\mathbin{/\mkern-6mu/}}
\newcommand{\ks}{{\bf k}_{\sslash}}
\newcommand{\ksd}{{\bf k}'_{\sslash}}
\newcommand{\rs}{{\bf R}_{\sslash}}
\newcommand{\rsd}{{\bf R}'_{\sslash}}
\newcommand{\rsdd}{{\bf R}''_{\sslash}}
\newcommand{\img}{\mathrm{i}}
\newcommand{\ha}{\mathscr{H}}

\newcolumntype{M}[1]{>{\centering\arraybackslash}m{#1}}

\begin{document}

\title{
Accurate multiscale simulation of frictional interfaces \\ by Quantum Mechanics/Green's Function molecular dynamics }

\author{Seiji Kajita}
    \email{fine-controller@mosk.tytlabs.co.jp}
    \affiliation{Toyota Central R\&D Labs., Inc., 41-1, Yokomichi, Nagakute, Aichi, 480-1192, Japan}
\author{Alberto Pacini}
    \affiliation{Department of Physics and Astronomy, University of Bologna, 40127 Bologna, Italy}
\author{Gabriele Losi}
    \affiliation{Department of Physics, Mathematics and Informatics, University of Modena and Reggio Emilia, 41125 Modena, Italy}
\author{Nobuaki Kikkawa}
    \affiliation{Toyota Central R\&D Labs., Inc., 41-1, Yokomichi, Nagakute, Aichi, 480-1192, Japan}
\author{Maria Clelia Righi}
    \email{clelia.righi@unibo.it}
    \affiliation{Department of Physics and Astronomy, University of Bologna, 40127 Bologna, Italy}

\begin{abstract}
Understanding frictional phenomena is a fascinating fundamental problem with huge potential impact on energy saving. Such an understanding requires monitoring what happens at the sliding buried interface, which is almost inaccessible by experiments. Simulations represent powerful tools in this context, yet a methodological step forward is needed to fully capture the multiscale nature of the frictional phenomena. Here, we present a multiscale approach based on linked \emph{ab initio} and Green's function molecular dynamics, which is above the state-of-the-art techniques used in computational tribology as it allows for a realistic description of both the interfacial chemistry and energy dissipation due to bulk phonons in non-equilibrium conditions. By considering a technologically-relevant system composed of two diamond surfaces with different degrees of passivation, we show that the presented method can be used not only for monitoring in real-time tribolochemical phenomena such as the tribologically-induced surface graphitization and passivation effects but also for estimating realistic friction coefficients. This opens the way to \emph{in silico} experiments of tribology to test materials to reduce friction prior to that in real labs.
\end{abstract}

%Use showkeys class option if keyword display desired
\maketitle

%-------------------------------
\section{Introduction}

 It is estimated that nearly one-third of the energy produced by fossil fuels to power vehicles is spent to overcome friction~\cite{Holmberg-2017}. Improved tribology technologies could dramatically reduce fuel consumption and CO\textsubscript{2} emissions. However, with respect to other technologies based on materials, tribology is remarkably less advanced. The reason resides in the complexity and variety of the phenomena that occur at the sliding buried interface, which is difficult to monitor in real-time by experiments. Simulations have a great potential to advance tribology, particularly those based on quantum mechanics, which is important for an accurate description of the chemical processes in conditions of enhanced reactivity. However, ab initio simulations as well as most of the atomistic methods nowadays used in tribology do not account for the energy dissipation by phonons. \\       
At the atomistic level, frictional forces appear during the relative motion of two surfaces in contact because their interaction energy changes as a function of the relative lateral position, giving rise to a corrugated potential energy surface (PES). The energy for climbing the PES hills, provided by the external force, is partially lost in non-adiabatic hill descents via phonon excitation. It is clear from this simplified description of the frictional slip that the amount of dissipated energy is governed by two main factors: the PES corrugation and the phonon propagation into the bulks in contact. The PES corrugation is determined by the electronic properties of the interface~\cite{Wolloch-2018}, while phonon excitation and propagation depend on the elastic properties of the infinite bulks. The latter also determines how the applied mechanical stresses are transferred to the sliding interface. \emph{In silico} experiments able to provide a quantitative estimate of the kinetic friction coefficient should then rely on a multiscale approach that includes both the electronic degrees of freedom at the interface and the vibrational degrees of freedom in the semi-infinite bulks.

Such a multiscale scheme is highly desirable also to accurately describe the activation mechanisms of tribochemical reactions, chemical processes involving environmental or lubricant molecules confined at the sliding buried interface.  The rate of these processes is highly accelerated with respect to reactions thermally activated at the open surface in static conditions~\cite{Hsu-2002,Zilibotti-2013}.
For example, thin films known as "tribofilms" are synthesized \emph{in situ} by mechanical rubbing additive molecules confined within micro-asperities contacts. These films are critical in preventing the cold sealing of nanoasperities and reduce the macroscopic friction and wear resistance of operating machinery parts.
Mechanosynthesis, which exploits impact forces to efficiently produce functional compounds and medicines without the use of solvents~\cite{Stuart-2012,Friscic-2013} is another important example where the control of the stress-assisted reactions is highly desirable.

Quantum-mechanics (QM) based molecular dynamics (MD) simulations can uncover elementary mechanisms of tribochemical/mechanochemical processes~\cite{Zilibotti-2013, nam-van-tran-2021}. Indeed, they have provided useful insight into several tribological phenomena such as the effects of humidity on the lubricity of carbon-based coatings,\cite{Kajita-2016, Moseler-2017} 2D materials like graphene, and transition metal dichalcogenides~\cite{Restuccia-2016, Levita-2015, Stella-2017}. Moreover, they allowed monitoring in real-time the first stages of tribofilm formation from commercial additives~\cite{Peeters-2020, NVT-2018, Kubo-2018} or hydrocarbon molecules~\cite{Giulio-Erdemir}. However, these simulations cannot be used to quantify the kinetic friction coefficient, because the limited thickness of the slabs which is typically used to model the solids in contact is too thin to contain the wavelength of the dissipated phonons. 
Indeed, several studies have reported that the energy dissipation associated with phonons, such as thermal conductivity and friction, are critically dependent on the size of the simulated systems~\cite{melis2014calculating, kajita2009deep, kajita2010approach, kajita2012simulation}.

Green's function (GF) molecular dynamics simulations and the related theory have been used to unleash the limitation of the limited system size~\cite{zwanzig1960collision, sokoloff1990theory, braun2005transition, campana2006practical, cai2001anisotropic, kajita2009deep, kajita2010approach, kajita2012simulation, kajita2016green, Monti-2021}.
This approach projects the dynamical response of all the degrees of freedom of the infinite solid atoms into a Green's function, which can excite phonons of any long wavelengths that propagate toward the infinite bulk system without reflection.  In other words, the phonon dissipation is implemented in a slab system, even though only the finite degrees of freedom are actually calculated.
Convolutions of the Green's function with applied forces represent effective forces of the surface atoms, that take the infinite solid atoms into account in the dynamics. However, the calculation of the convolution is a critical computational bottleneck for the use of GF MD. A solution for such a problem has been recently proposed for general surfaces~\cite{kajita2016green}, based on the elegant analytical solution of the Green's function using a fast convolution method~\cite{talbot1979accurate, lubich2002fast,capobianco2007fast,phdPrete}. Therefore, the GF MD method can now be applied to large-scale simulations of realistic systems previously considered too computationally demanding.
However, this framework is based on classical force fields, where the electronic degrees of freedom necessary to accurately describe the surface-surface interaction and the tribochemical processes are not considered.

To overcome this limitation, we propose a new multiscale approach that combines the strengths of the QM MD and GF MD. 
This is realized by a hybrid method that links the quantum-mechanical and GF molecular-mechanical parts of the system \cite{swart2003addremove}.
The hybrid QMGF MD method can be used to obtain accurate quantitative estimates of the friction forces taking both interface chemistry and phonon dissipation into account. This can open the way to a novel understanding of tribological phenomena and allows for the execution of accurate tribochemistry experiments \emph{in silico}.

The manuscript is organized as follows: section \ref{sec:methods} presents the theoretical framework (\ref{sec:1d-GF}-\ref{sec:methodsgs}) and the computational implementation (\ref{sec:methodsmd}-\ref{sec:compdetails}) of the GF MD method. An example of application is provided in section \ref{sec:results}, focusing on the tribological properties of diamond as a function of surface hydrogenation and showing that the hybrid QMGF MD method is able to provide a quantitative estimation of the kinetic friction coefficients in agreement with experiments. Finally the conclusions of this work are given in section \ref{sec:conclusion}.

\section{Methods} \label{sec:methods}

Here we describe the theoretical method and the numerical strategies implemented in the developed multiscale code. We start by reviewing the GF MD methodology shown in our previous work~\cite{kajita2016green} and then include details on the fast convolution and thermo-barostats.
Finally, the hybrid add-remove method is presented, along with a numerical strategies for stabilizing the dynamics in QMGF MD simulations.

\subsection{Green's function of a one-dimensional chain}\label{sec:1d-GF}
We begin with a semi-infinite one-dimensional chain as a simple example, which assists the readers in understanding the GF MD for general surfaces presented later.
A chain composed of harmonic oscillators is considered, and
the atoms are identical and connected with monotonic bonds modeled with springs of constant $K$. 
The equations of motion are:
\begin{eqnarray}
m \frac{d^2}{dt^2} u_1(t) &=& -K(u_1 - u_2) + f_1(t) \ \ \ \ \ i = 1 \nm \\
m \frac{d^2}{dt^2} u_i (t) &=& -K(2 u_i - u_{i+1}- u_{i-1}) \ \ \ \ \ 1 < i < \infty,  \nm 
\end{eqnarray}
where $m$ and $u_i$ are mass and displacement of the $i$th atom, respectively.
The external force $f_1$ is applied only on the edge atom $i=1$. 
The displacement of the edge atom is mathematically written as the convolution form of the Green's function $G$~\cite{StanleyJFarlow} with the force as
\begin{eqnarray}
 u_1(t) = \int_{0}^{t} G(t-\tau) f_1(\tau) d\tau. \label{eq:one-chain g0}
\end{eqnarray}
The Laplace transformation of Eq.~\ref{eq:one-chain g0} is
\begin{eqnarray}
 u_1(z) = G(z) f_1(z) \label{eq:one-chain g1},
\end{eqnarray}
where $z$ is a coordinate in the complex space.
It should be noted that $G(z)$ is numerically more important than $G(t)$ in GF MD with respect to both of the fast convolution and the thermo-barostats methods explained later.

An atom, labeled by $i=0$, is then coupled on top of the edge atom $i=1$. 
This new atom becomes a surface atom under an external force, and its equation of motion after the Laplace transformation is
\begin{eqnarray}
m z^2 u_0(z) &=& -K(u_0(z) - u_1(z)) + f(z), \label{eq:one-chain g2} 
\end{eqnarray}
where $f$ is an external force on the new edge atom $i=0$. 
Because
$f_1$ in Eq.~\ref{eq:one-chain g1} becomes counteracting force of the fist term in the right side of Eq. \ref{eq:one-chain g2},
Eq.~\ref{eq:one-chain g1} becomes
\begin{eqnarray}
 u_1(z) = K G(z) (u_0(z) - u_1(z))  \label{eq:one-chain g3}.
\end{eqnarray}
By inserting Eq.~\ref{eq:one-chain g3} into Eq.~\ref{eq:one-chain g2} to eliminate $u_1$, we obtain
\begin{eqnarray}
 u_0(z) = \left( mz^2 + \frac{K}{1+KG(z)} \right)^{-1} f(z)  \label{eq:one-chain g4}.
\end{eqnarray}

An important argument is that this addition of the $i=0$ atom to a semi-infinite system in this way does not essentially change the original system due to its infinity.
This invariant feature, called semi-infinite periodicity, simplifies the derivation of the Green's function. 
Namely, because the periodicity tells that Eq.~\ref{eq:one-chain g1} is equivalent to Eq.~\ref{eq:one-chain g4}, we can derive
\begin{eqnarray}
 G(z) = \left( mz^2 + \frac{K}{1+KG(z)} \right)^{-1}  \label{eq:one-chain g5}.
\end{eqnarray}
Equation~\ref{eq:one-chain g5} is readily solved as
\begin{eqnarray}
 G(z) = \frac{1}{2K} \left( - 1 + \sqrt{1+\frac{4K}{mz^2}} \right). \nm % \label{eq:one-chain g5}.
\end{eqnarray}
Derivation of $G(z)$ without using the semi-infinite periodicity is more complicated as shown in Ref.~\cite{kajita2010approach}.
Semi-infinite periodicity is a key to generalize the method applicable to any surface system.

\subsection{Green's function of a general surface} \label{sec:methodsgs}
%---------------------

\begin{figure}[htbp]
\centering
\includegraphics[width=0.7\linewidth]{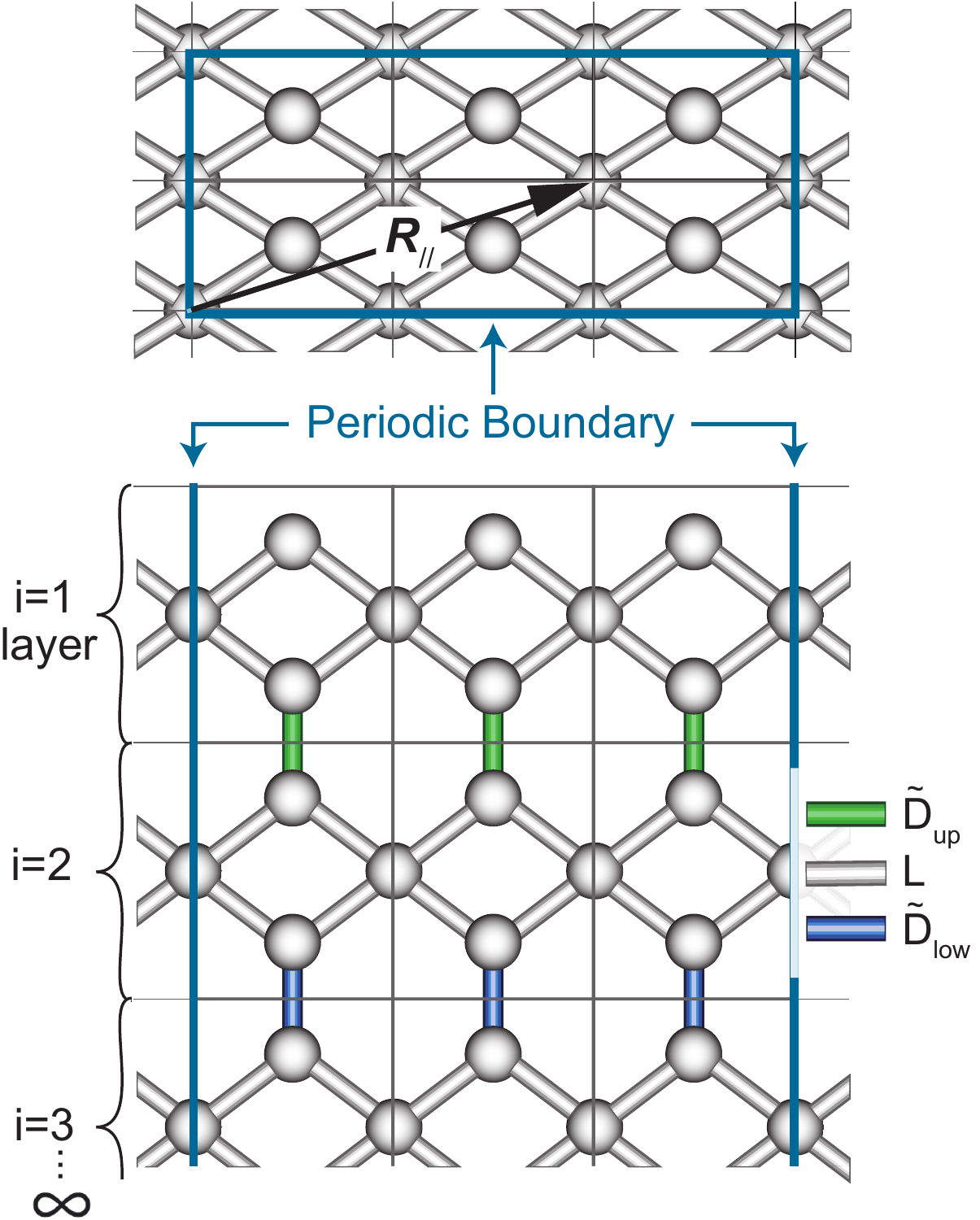} 
\caption{
Schematic image of the semi-infinite surface system. 
The supercell consists of unit cells, and it repeats periodically along the surface lateral direction and infinitely along the direction normal to the surface. The layers are labeled by indices $i$ that increase as going to the bulk direction. The colored bonds indicate inter-layer terms  ${\tilde D}_{{\textrm up}}$ and ${\tilde D}_{{\textrm low}}$ of the $i=2$ layer; these quantities are used in  Sec \ref{sec:methodsgs}.
}
\label{pic:model}
\end{figure}

%---------------------

The strategy to derive the Green's function of a general three-dimensional semi-infinite solid is the same as the one-dimensional chain.
Let us consider a general crystalline surface as shown in Fig.~\ref{pic:model}.
We define a surface layer that is a set of unit cells laterally aligned in the periodic boundary conditions. Each layer is labeled with an index starting from the surface $i=1,2,\cdots \infty$.
This concatenation of the layers in the surface normal direction constitutes the semi-infinite solid.
We write the equation of motion for the system as
\begin{eqnarray}
\left( M \frac{d^2}{dt^2} + D \right) {\bf u}(t) = {\bf f}_1(t), \label{eq:me0}
\end{eqnarray}
where $M$ is an atomic-mass diagonal matrix.
The vector ${\bf u}$ represents the atomic displacements of the entire system, where ${\bf u} = ({\bf u}_1, {\bf u}_2, \cdots)^T$ and ${\bf u}_i$ is the displacement vector of the atoms in the $i$th layer.
The external force vector ${\bf f}_1$ is applied only to the surface layer $i=1$.
The $D$ matrix is referred to as an internal-force matrix that represents elastic constants of bonds for all the atoms.
Vectors, matrices, and scalars are indicated in bold, uppercase and lowercase letters, respectively.
We normalize Eq.~\ref{eq:me0} by the mass using a $N = M^{-1/2}$ operator.
\begin{eqnarray}
\left( \frac{d^2}{dt^2} + {\tilde D} \right)  {\tilde {\bf u}}(t) = {\tilde {\bf f}}_1(t), \label{eq:me1}
\end{eqnarray}
where we define ${\tilde D} = N D N,  {\tilde {\bf u}} = N^{-1} {\bf u}$ and ${\tilde {\bf f}}_1 = N {\bf f}_1$.

A standard approach to include the periodic boundaries is the discrete Fourier transformation.
A set of surface lattice vectors $\rs$ points to the origins of the lateral positions of the constituent unit cells in the layer (see the upper panel of Fig.~\ref{pic:model}).
The discrete Fourier transformation of arbitrary vector ${\bf x}$ and matrix $X$ of the layer are
\begin{eqnarray}
{\bf x}(\ks) &=& \sum_{\rs} \exp(-\mathrm{i} \ks\cdot \rs) 
{\bf  x}(\rs)/\sqrt{n}
, \nm \\
X(\ks, \ksd)&=& \sum_{\rs, \rsd} 
\exp(-\mathrm{i} \ks \cdot \rs) X(\rs, \rsd) \nm \\
  & &\ \ \ \ \ \ \ \ \ \times \exp(\mathrm{i} \ksd \cdot \rsd)/n,  \nm
\end{eqnarray}
where $\ks$ is the  surface reciprocal vector of $\rs$, and $n$ is the number of unit cells in the layer. 
We use a matrix notation $X(\ks)$ when the matrix is diagonal with the $\ks$ basis.

According to Bloch's theorem, the internal-force matrix $D$ is diagonal in the $\ks$ basis due to the inherent periodicity of the system.
In the initial conditions ${\bf u}(t=0)={\dot {\bf u}}(t=0) = 0$, Eq.~\ref{eq:me1}
becomes
\begin{eqnarray}
{\tilde {\bf u}}(z, \ks) = G (z, \ks) {\tilde {\bf f}}_1(z,\ks), \nonumber \\
 G(z, \ks)=\left( z^2 + {\tilde D}(\ks) \right)^{-1}, \label{eq:pureG} 
\end{eqnarray}
after the discrete Fourier and Laplace transformations.
Recalling the external force vector ${\tilde {\bf f}}_1$ is applied only on the surface layer $i=1$, ${\tilde {\bf u}}_1$ is 
\begin{eqnarray}
{\tilde {\bf u}}_1(z, \ks) = G_{11} (z, \ks) {\tilde {\bf f}}_1(z,\ks), \label{eq:gk}
\end{eqnarray}
where $G_{ii'}$ is the corresponding element of $G$ in the layer indices $i$ and $i'$. 
In the $\rs$ basis, Eq.~\ref{eq:gk} becomes
\begin{eqnarray}
{\tilde {\bf u}}_1(z, \rs) = \sum_{\rsd} G_{11}(z, \rs - \rsd) 
{\tilde {\bf f}}_1(z, \rsd). \label{eq:geta} 
\end{eqnarray}

An additional layer $i=0$ is piled up on the surface system by connection with the $i=1$ layer.
Before applying the semi-infinite periodicity, we decompose the internal-force matrix ${\tilde D}$ into an intra-layer term $L$ that represents the bonds within the layer, and inter-layer terms 
${\tilde D}_{{\textrm low}} \oplus {\tilde D}'_{{\textrm low}}$ and ${\tilde D}_{{\textrm up}} \oplus {\tilde D}'_{{\textrm up}}$, representing the bonds to the lower and upper layers, respectively (see Fig.~\ref{pic:model}).
Namely, the matrix representation of ${\tilde D}$ in the layer index is
\begin{eqnarray}
{\tilde D} &=& 
 \begin{pmatrix}
  L       & 0       & 0      & 0      & \cdots  \\
  0       & L       & 0      & 0      & \cdots  \\
  0       & 0       & L      & 0      & \cdots  \\
  \vdots  & \vdots  & \vdots & \ddots & \vdots
 \end{pmatrix}
 \nonumber \\
 &+&
 \begin{pmatrix}
  {\tilde D}_{{\textrm low}}  & {\tilde D}'_{{\textrm low}}                  & 0                                  & 0                   & \cdots  \\
  {\tilde D}'_{{\textrm up}}  & {\tilde D}_{{\textrm up}} + {\tilde D}_{{\textrm low}} & {\tilde D}'_{{\textrm low}}                  & 0                   & \cdots  \\
  0                 & {\tilde D}'_{{\textrm up}}                   & {\tilde D}_{{\textrm up}} + {\tilde D}_{{\textrm low}} & {\tilde D}'_{{\textrm low}}   & \cdots  \\
  \vdots            & \vdots                             & \vdots                             & \ddots              & \vdots 
 \end{pmatrix}
\nm .
\end{eqnarray}

As in the previous subsection, we first write the equation of motion of the new layer
\begin{eqnarray}
z^2 {\tilde {\bf u}}_0(z, \rs) = &-& \sum_{\rsd} L(\rs - \rsd)  {\tilde {\bf u}}_0(z, \rsd) \nm\\
                                 &-& {\tilde D}_{{\textrm low}} {\tilde {\bf u_0}}(z, \rs) \nm \\
                                 &-& \sum_{\rsd} {\tilde D}'_{{\textrm low}}(\rs - \rsd) {\tilde {\bf u}}_1(z, \rsd) \nm \\
                                 &+& {\tilde {\bf f}}(z, \rs), \label{eq:simul1}
\end{eqnarray}
The external force ${\tilde {\bf f}}$ is  applied only on the $i=0$ layer. 
Then, giving that ${\tilde {\bf f}}_1$ is a counteracting elastic force between ${\tilde {\bf u}}_0$ and ${\tilde {\bf u}}_1$,
 Eq.~\ref{eq:geta} becomes
\begin{eqnarray}
& &{\tilde {\bf u}}_1(z, \rs) = \sum_{\rsd} G_{11} (z, \rs - \rsd)  \nonumber \\ 
&\times& 
\left(
{\tilde D}_{{\textrm up}} {\tilde {\bf u}}_1(z, \rsd) \!+\! \sum_{\rsdd} {\tilde D}'_{{\textrm up}}(\rsd - \rsdd) {\tilde {\bf u}}_0 (z, \rsdd)
\right). \nonumber \\
\label{eq:simul2}
\end{eqnarray}
The Fourier transformations of Eqs.~\ref{eq:simul1} and \ref{eq:simul2}  yield
\begin{eqnarray}
z^2  {\tilde {\bf u}}_0(z, \ks) = 
 &-& \left( L(\ks) +  {\tilde D}_{{\textrm low}}  \right)  {\tilde {\bf u}}_0(z, \ks) \nm \\
 &-& {\tilde D}'_{{\textrm low}}(\ks) {\tilde {\bf u}}_1(z, \ks)  + {\tilde {\bf f}}(z, \ks),
 \label{eq:simul2-1}
\end{eqnarray}
\begin{eqnarray}
 {\tilde {\bf u}}_1(z, \rs) &=&  G_{11} (z, \ks) \nm \\
 &\times& \left( {\tilde D}_{{\textrm up}} {\tilde {\bf u}}_1(z, \ks) + {\tilde D}'_{{\textrm up}}(\ks) {\tilde {\bf u}}_0 (z, \ks) \right) .
\label{eq:simul2-2}
\end{eqnarray}
Inserting Eq.~\ref{eq:simul2-2} into Eq.~\ref{eq:simul2-1} to erase ${\tilde {\bf u}}_1$, we obtain,
\begin{eqnarray}
{\tilde {\bf u}}_0(z, \ks) =  \left(z^2 + L(\ks) + {\tilde D}^{*}(z, \ks) \right)^{-1} {\tilde {\bf f}}(z, \ks), \label{eq:simul3}
\end{eqnarray}
where ${\tilde D}^{*}$ is an effective interlayer matrix defined by
%\begin{widetext}
\begin{eqnarray}
{\tilde D}^{*} = {\tilde D}_{{\textrm low}}  
&+& {\tilde D}'_{{\textrm low}}(\ks) \left(1 - G_{11}(z, \ks)\ {\tilde D}_{{\textrm up}}\right)^{-1} \nm \\
& & \times G_{11}(z, \ks)\
{\tilde D}'_{{\textrm up}}(\ks).
\nm  %\label{eq:effectiveK}
\end{eqnarray}
%\end{widetext}

Finally, the semi-infinite periodicity promises that Eqs.~\ref{eq:gk} and \ref{eq:simul3} are equivalent because the new layer should respond to external forces in the entirely same manner as the original surface. 
We obtain an equation for the Green's function as
\begin{eqnarray}
G_{11}(z, \ks) = \left(z^2 + L(\ks) + {\tilde D}^{*}(G_{11}) \right)^{-1}.\label{eq:gmain}
\end{eqnarray}
The matrices $L$ and ${\tilde D}$ are numerically estimated by phonon calculations of the bulk system based on {\it ab initio} calculations.
Equation ~\ref{eq:gmain} is solved by conventional Newton-Raphson algorithms.
%The trajectory of the mass-normalized surface layer ${\tilde {\bf u}}_1$ 
%can be obtained by a convolution of the Green's function as
%\begin{eqnarray}
%{\tilde {\bf u}}_1(t, \ks) = \int^{t}_{0} G_{11} (t-t', \ks) {\tilde {\bf f}}(t',\ks) dt',  %\label{eq:gkt}
%\end{eqnarray}
%in the initial conditions ${\bf u}(t=0)={\dot {\bf u}}(t=0) = 0$.

%%---------------------------------------------------------------
\subsection{Green's function molecular dynamics} \label{sec:methodsmd}
The three-dimensional displacements of the semi-infinite surface layer atoms are described by a linear combination as:
\begin{eqnarray}
{\bf u}_p + {\bf u}_g, 
\label{eq:upug}
\end{eqnarray}
where ${\bf u}_p, {\bf u}_g \in {\bf R}^{N\times3}$ are 
a particular solution and general solution 
~\cite{StanleyJFarlow} of the equation of motion, respectively; and $N$ is the number of the surface atoms in the unit cell. 
The solution ${\bf u}_p$ represents trajectories driven by an external force ${\bf f}$ applied on the surface layer, at initial conditions ${\bf u}_p(t=0) = 0, \frac{d}{dt} {\bf u}_p(t=0) = 0$.\newline
The general solution ${\bf u}_g$, on the other hand, is that without external force but in arbitrary initial conditions ${\bf u}_g(t=0), \frac{d}{dt} {\bf u}_g(t=0)$.
Notably, ${\bf u}_g$ can represent the thermostat and barostat of the system
when their statistic features are related to the Green's function.\newline
This subsection provides numerical recipes on how to compute ${\bf u}_p $ and ${\bf u}_g $.
 
\subsubsection{Particular solution and convolution}
%The modification of Eq.~\ref{eq:gkt} gives
By using the Green's function in Eq. \ref{eq:gmain},  
the equation of motion of ${\bf u}_p$ can be written as 
\begin{eqnarray}
M \frac{d^2}{dt^2} {\bf u}_p(t, \ks) &=& {\bf f}_{\textrm{GF}}(t,\ks) \label{eq:psol}\\
{\bf f}_{\textrm{GF}}(t,\ks) &=& \int_0^{t} A(t-\tau,\ks) {\bf f}(\tau,\ks) d\tau \label{eq:reducedF}, %\\
%A(t,\ks) &=& N \frac{d^2}{dt^2} G_{11}(t,\ks) N, \nm
\end{eqnarray}
where the Laplace transformed $A$ is defined as $A(z, \ks) = z^2 N^{-1} G_{11}(z,\ks) N$
and ${\bf f}$ is an applied force on the surface layer.
The reduced force ${\bf f}_{\textrm{GF}}$ has a convolution form, which becomes a computational bottleneck if discrete integral algorithms are used.
The integral range grows as time $t$ increases and the entire history of the force trajectory should be saved in memory. 
Indeed, the simulation time and memory allocation are proportional to $O(t^2)$ and $O(t)$, respectively.

%---------------------

\begin{figure}[htbp]
%\centering
\includegraphics[width=1.0\linewidth]{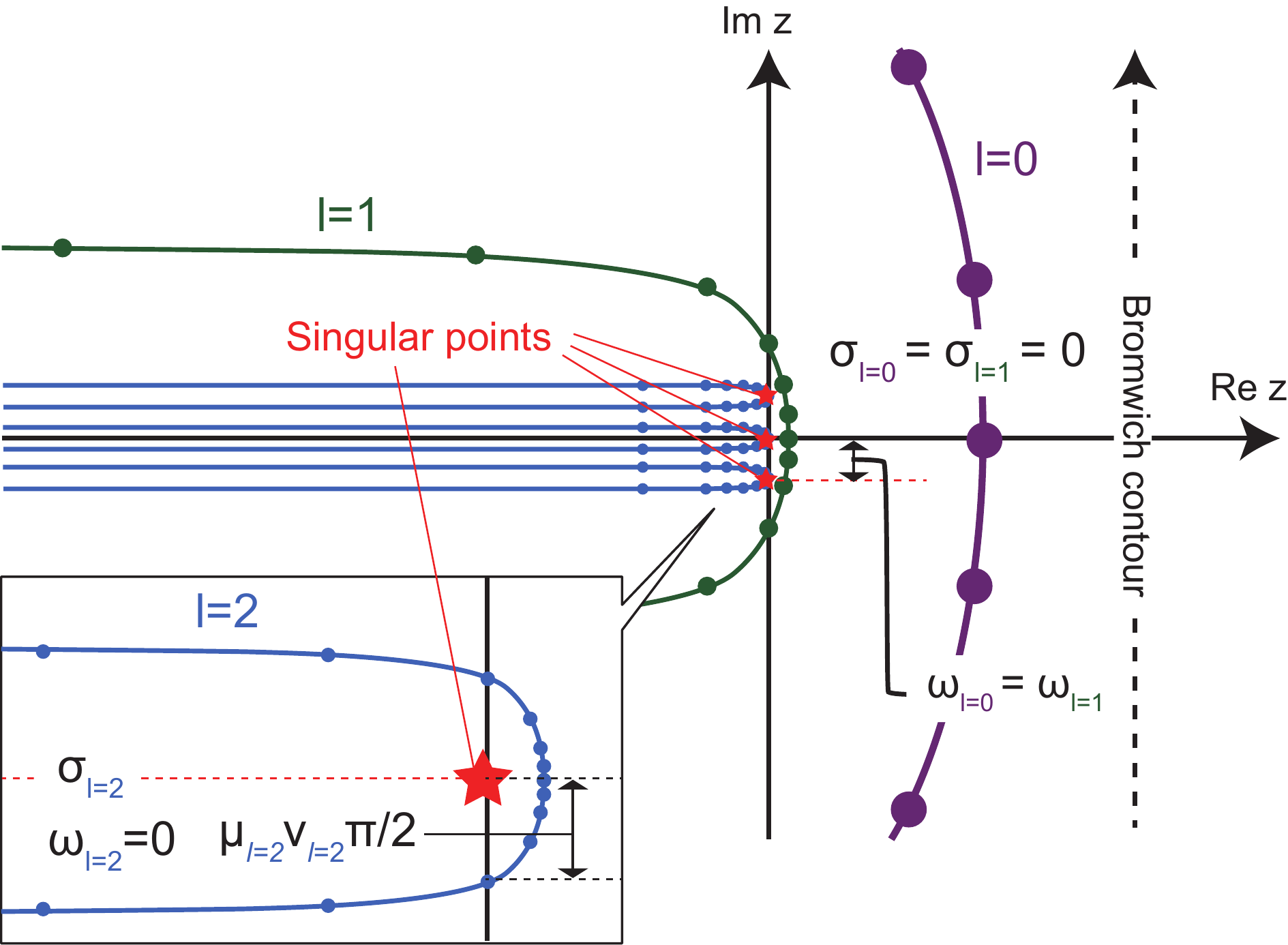} 
\caption{
Schematic of integral paths of the mTILT.
The paths labeled by index $l$ are used in the time range $I_l$ in Eq. \ref{eq:interval}.
The parameter $\sigma_l$ indicates a center of the contour in the imaginary axis, which is set with respect to imaginary parts of the singular points.
The $\omega_l$ is assigned to the distance of the imaginary parts between the contour center $\sigma_l$ and the farthest singular point.
The width of the $l$ path is the distance of the imaginary parts between $\sigma_l$ and a cross-section of the contour with the imaginary axis, which is equivalent to $\mu_l \nu_l \pi/2$.
}
\label{pic:integralPath}
\end{figure}

%---------------------
A fast convolution based on modified-Talbot's inverse Laplace transformation (mTILT)\cite{talbot1979accurate, lubich2002fast,capobianco2007fast,phdPrete} reduces this notorious
computational costs into $O(t\log(t))$ for the simulation time and $O(\log(t))$ for memory allocation.
The conventional inverse Laplace transformation of an arbitrary function $X(z)$ is defined by
\begin{eqnarray}
X(t) = \frac{1}{2\pi \img}\int_{c-\img\infty}^{c+\img\infty} X(z) e^{zt} dz, \nm
\end{eqnarray}
where the constant $c$ is a real number larger than zero. This integral path is called Bromwich contour. The idea of mTILT is that the Bromwich contour is bent in such a way as to encircle singular points of $X(z)$ on the imaginary axis, as shown in Fig.~\ref{pic:integralPath}.
Coordinates of the singular points of the Green's function are identified by a line search of $G_{11}(z=\img \omega', \ks)$, where $\omega'$ is a real number variable.
The mTILT divides the time range [0,$T$] into a set of time ranges $I_l$ as follows
\begin{eqnarray}
I_0 = [0,h], \ I_{l} = [B^{l-1}h, T_l ], \ T_l = (2B^l-1)h, \label{eq:interval}
\end{eqnarray}
where $h$ is a time step and $l = 1, 2, \cdots, L$. The integer $L$ satisfies $(2B^{L}-1)h \ge T$ and $B$ is an arbitrary integer greater than $1$. The integral path used in $I_l$ is defined as
\begin{eqnarray}
z^{l}(\theta) = \img \sigma_l + \mu_l(\theta \cot(\theta) + \img \nu_l\theta), \nm
\end{eqnarray}
where the geometry parameters are $\mu_l = \mu_0/T_l$, $\mu_0 =8$, 
$\nu_l = \nu_0(1+\omega_l/\beta)$, $\nu_0=0.6$, and 
$\beta = \pi \mu_l \nu_0 /2$.

For example, Fig.~\ref{pic:integralPath} illustrates shapes of the paths $z^l$.
The widths of $l=0$ and $l=1$ paths are large enough to enclose all of the three singular points. 
%The centers of the contours are $\sigma_{l\le1}=0$, while $\omega_{l\le1}$ is set to the farthest singular point.
As $l$ increases, the width of the path $\mu_l \nu_l \pi/2$ decreases because $\mu_l$ decreases.
The number of $l=2$ paths, in this example, becomes three, and each path encloses each of the singular points.
%The three contours have different $\sigma_l$, which corresponds to the imaginary coordinate of their assigned singular point.
%The parameter $\omega_l$ is zero because of the integral circuit of the single singular point.

In this manner, the mTILT designs the integral paths to secure high numerical accuracy of the inverse Laplace transformation, depending on the time interval $I_l$. Namely, when $t \in I_l$, the inverse Laplace transformation is
\begin{eqnarray}
X(t) &=& \frac{1}{2\pi \img}\int_{-\pi}^{\pi} X(z^l(\theta)) e^{z^l(\theta)t} \frac{dz^l}{d\theta} d\theta, \nm \\
      &\sim& \frac{1}{2 \img (N+1)} \sum_{j=-N}^{N} X(z^{l}(\theta_j)) e^{z^{l}(\theta_j)t} \frac{dz^{l}(\theta_j)}{d\theta} \nm \\
      &\equiv&  \sum_{j=-N}^{N}  \omega^l_j  X(z^{l}_j) e^{z^{l}_j t}, \label{eq:mtilt}
\end{eqnarray}
where a trapezoidal rule in the integral range $[-\pi, \pi]$ is used with discretization $\theta_j = j \pi / N +1, j = -(N+1), \cdots, N+1$.
We defined $\omega^l_j = \frac{d}{d\theta} z^l_j  /2 \img (N+1)$ and $z^l_j = z^l(\theta_j)$.
For notation simplicity, Eq.~\ref{eq:mtilt}, which represents the single path embracing all of the singularity, will be used in the following. In the case of the plural paths as $l=2$ in Fig.~\ref{pic:integralPath}, contributions calculated by Eq.~\ref{eq:mtilt} are merely summed up.

Then, the mTILT is applied to the convolution task in Eq.~\ref{eq:reducedF}. 
A range of the simulation time $[0,T]$ is divided according to Eq.~\ref{eq:interval} in the convolution routine.
Namely, when $[t-a, t-b] \subseteqq I_l$, the convolution is approximated as
\begin{eqnarray}
&&\int_a^{b} A(t-\tau,\ks) {\bf f}(\tau,\ks) d\tau \nm \\
&&\sim \sum_{j=-N}^{N} \omega^l_j  A(z^{l}_j) e^{z^{l}_j (t-b)}  {\bf y}(b, a, z^l_j, \ks),  \label{eq:iltapprox0}
\end{eqnarray}
where ${\bf y}(b, a, z,\ks) = \int^b_a e^{z (b-\tau)} {\bf f}(\tau,\ks) d\tau$.
Here we omit $\ks$ variable unless the context needs it explicitly.
The quantity ${\bf y}(b, a, z)$ is known to be a solution of the following differential equation at $t=b$,
\begin{eqnarray}
\frac{d}{dt} {\bf y}(t, a, z) = z {\bf y}(t, a, z) + {\bf f}(t), \ \ \ {\bf y}(a, a, z) = 0. \label{eq:eqZ0}
\end{eqnarray}
We then approximate ${\bf y}$ by time-discretized ${\bf f}(t_k)$.
The time interval $[a, b]$ is split into a sequence of partial intervals $[a+t_k, a+t_{k+1}]$, where $t_k = k \times h$ and $k = 0, \cdots, n=(b-a)/h$.

In $t \in [a+t_k, a+t_{k+1}] \subset I_l$, Eq.~\ref{eq:eqZ0} is expressed as
\begin{eqnarray}
\frac{d}{dt} {\bf y}(t, a, z) &=& z {\bf y}(t, a, z) + {\bf f}(t), \nm \\ 
{\bf y}(a+t_k, a, z) &=& {\bf y}_k. \nm %\label{eq:eqZ1}
\end{eqnarray}
An exact solution of this equation is 
 \begin{eqnarray}
{\bf y}_{k+1} = e^{zh} {\bf y}_k + h \int_0^{1} e^{(1-\theta)zh} {\bf f}(a+t_k+h\theta) d\theta. \nm
\end{eqnarray}
We apply a linear approximation ${\bf f}(a+t_k+h\theta) \sim \theta {\bf f}_{k+1} + (1-\theta) {\bf f}_k$, 
where $ {\bf f}_k =  {\bf f}(a+t_k)$.
As a result, the approximated solution ${\bf y}'_{k}$ can be obtained via a recursive expression with respect to the index $k$,
\begin{eqnarray}
{\bf y}'_{k+1} &=& {\bf y}'_{k} + 
                        \frac{e^{zh}-1}{zh} \left( zh{\bf y}'_{k} + h {\bf f}_k + h \frac{{\bf f}_{k+1} - {\bf f}_k}{zh}  \right) \nm \\
                        && - h \frac{{\bf f}_{k+1} - {\bf f}_k}{zh}. \label{eq:zapprox}
\end{eqnarray}

Since the mathematical components have been prepared, we now describe the fast convolution integral. Denoting $t=t_{n+1}$, we divide the range of convolution into two regions as
\begin{eqnarray}
\left(
\int_{t_n}^{t_{n+1}} +  \int_0^{t_n} 
\right)
A(t_{n+1}-\tau) {\bf f}(\tau) d\tau. 
\label{eq:convol-tmp}
\end{eqnarray}
The modified-Talbot path of $I_0$ calculates the first term as
\begin{eqnarray}
&& \int_{t_n}^{t_{n+1}} A(t_{n+1}-\tau) {\bf f}(\tau) d\tau  \nm \\
&& \sim \Phi_1  {\bf f}(t_n)  + \Phi_2 \frac{{\bf f}(t_{n+1}) - {\bf f}(t_n)}{h}, \label{eq:convol1}
\end{eqnarray}
where
\begin{eqnarray}
\Phi_1 &=& \int_{t_n}^{t_{n+1}} A(t_{n+1} - \tau) d\tau  \nm \\ 
    &=& \int_{0}^{h}  A(h - \tau) d\tau 
   \sim \sum_{j=-N}^{N} \omega^0_j \frac{A(z^0_j)}{z^0_j}e^{z^0_j h} \nm \\
\Phi_2 &=& \int_{t_n}^{t_{n+1}} A(t_{n+1} - \tau) (t_{n+1} - \tau) d\tau \nm \\ 
   &=& \int_{0}^{h}  A(h - \tau) \tau d\tau 
   \sim \sum_{j=-N}^{N} \omega^0_j \frac{A(z^0_j)}{(z^0_j)^2}e^{z^0_j h}. \nm 
\end{eqnarray}
The second term of Eq.~\ref{eq:convol-tmp} is decomposed into contributions of the intervals $I_{l=1,2, \cdots, L-1}$, where $L$ is the minimum integer that satisfies $t_{n+1} < 2B^{L}h$.
We define $\tau_0 = t_n$, $\tau_L = 0$, and $\tau_l = q_l B^{L} h$ if $l\neq 0$ nor $L$.
An integer $q_l \ge 1$ is determined so as to satisfy
\begin{eqnarray}
t_{n+1} - \tau_l \in [B^l h, (2B^l -1)h], \ l=1,2, \cdots, L-1. \nm
\end{eqnarray}
The time range is divided as $[0, t_n] = \cup_{l=0} [\tau_l, \tau_{l-1}]$.
Therefore, by using the approximations in Eqs.~\ref{eq:iltapprox0} and \ref{eq:zapprox}, we derive 
\begin{eqnarray}
&&\int_{0}^{t_n} A(t_{n+1}-\tau) {\bf f}(\tau) d\tau \nm \\
&& = \sum_{l=1}^{L} \int_{\tau_l}^{\tau_{l-1}}  A(t_{n+1}-\tau) {\bf f}(\tau) d\tau \nm \\
&& \sim \sum_{l=1}^{L} \sum_{j=-N}^{N} \omega^l_j  A(z^{l}_j) e^{z^{l}_j (t_{n+1}-\tau_{l-1})}  {\bf y}'(\tau_{l-1}, \tau_l, z^l_j). 
\label{eq:convol2}
\end{eqnarray}
In short, the convolution term is calculated by the sum of Eqs.~\ref{eq:convol1} and \ref{eq:convol2}, along with Eq.~\ref{eq:zapprox} which is used for efficient calculation of the term ${\bf y}'$.
Then, because the reduced force is obtained, the motion equation Eq.~\ref{eq:psol} is numerically solved to simulate the trajectory.
A simple example of specific steps to show the ${\bf y}'$ updating is given in the Supplementary Information.

\subsubsection{General solution and thermo-barostats}\label{subsec:thermo_baro}
We consider the general solution ${\bf u}_g$ in Eq.~\ref{eq:upug}. 
Let us use ${\bf u}$ instead of  ${\bf u}_g$ because of notation simplicity.
The mass-normalized equation of motion for the {\it whole} system in the $\ks$ space is 
\begin{eqnarray}
\left( \frac{d^2}{dt^2} + {\tilde D}(\ks) \right)  {\tilde {\bf u}}(t, \ks)  = 0,  \label{eq:meg1} \\
{\tilde {\bf u}}(0, \ks) \equiv  {\tilde {\bf u}}(t=0, \ks), \nm \\
{\tilde {\bf v}}(0, \ks) \equiv \frac{d}{dt} {\tilde {\bf u}}(t=0, \ks), \nm
\end{eqnarray}
where ${\tilde {\bf u}} = ({\tilde {\bf u}_1}, {\tilde {\bf u}_2}, \cdots)^T$ in the layer index representation.
The Laplace transformation of Eq.~\ref{eq:meg1} yields
\begin{eqnarray}
\left( z^2 + {\tilde D}(\ks) \right)  {\tilde {\bf u}}(z, \ks)  = {\tilde {\bf v}}(0,\ks)  + z  {\tilde {\bf u}}(0,\ks). \nm
\end{eqnarray}
Using Eq.~\ref{eq:pureG}, we can describe the general solution in the Green's function framework, as 
\begin{eqnarray}
  {\tilde {\bf u}}(z, \ks)  = G(z, \ks) \left(
    {\tilde {\bf v}}(0,\ks)  + z  {\tilde {\bf u}}(0,\ks)
    \right).  \label{eq:meg2}
\end{eqnarray}

The initial condition includes all the displacements and velocities in the semi-infinite system. Obviously, there is an infinitely large number of possible configurations of the initial conditions.
A reasonable policy to select a physically meaningful one is to consider a thermostat. The semi-infinite system is assumed to be located at a temperature $T$, and the constituent atoms move according to the thermal fluctuation.
Let this general solution be denoted by ${\tilde {\bf u}}_{T}$.
We modify Eq.~\ref{eq:meg2} by using notations $S(z) \equiv zG(z)$ and ${\tilde {\bf v}}_T(z, \ks) \equiv z {\tilde {\bf u}}_T(z, \ks)$, and apply the inverse Laplace transformation.
 \begin{eqnarray}
  {\tilde {\bf v}}_T(t, \ks)  = S(t, \ks) \left(
    {\tilde {\bf v}}_T(0,\ks)  + z  {\tilde {\bf u}}_T(0,\ks)
    \right).  \label{eq:meg3}
\end{eqnarray}
Here we use the law of equipartition of energy:
 \begin{eqnarray}
  \langle {\tilde v}_{T; \xi}(t,\ks)   {\tilde v}^*_{T; \xi'}(t,\ks) \rangle  &=& k_B T \delta_{\xi, \xi'} \nm \\
  \langle {\tilde v}_{T; \xi}(t,\ks)   {\tilde u}^*_{T; \xi'}(t,\ks)  \rangle &=& 0, \nm
\end{eqnarray}
where $\xi$ refers to the components of the atomic coordinates (x, y, z), and $*$ is the complex conjugate. The bracket represents the ensemble averaging operator.
By applying the equipartition law, Eq.~\ref{eq:meg3} becomes
\begin{eqnarray}
\langle {\tilde v}_{T; \xi}(t, \ks)   {\tilde v}^*_{T; \xi'}(0,\ks) \rangle  &=& k_B T S_{\xi, \xi'}(t, \ks), \label{eq:temperature}
\end{eqnarray}
that is called the fluctuation-dissipation theorem.
This relation tells that an auto-correlation of the general solution of the velocity should be equivalent to the Green's function.

% \subsubsection{General solution: stress}
Another useful general solution represents normal and shear stresses.
A semi-infinite system is located at 0~K temperature under a uniform stress applied to the surface ${\bf f}(t, \ks) = - \delta_{\ks, {\bf 0}} {\bf f}_{s}$.
The velocity solution of this system is
\begin{eqnarray}
{\tilde {\bf v}}(t, \ks)  = - \delta_{\ks, {\bf 0}} {\bf f}_{s} \int^t_0 S_{11}(\tau, \ks) d\tau,  \nm %\label{eq:stress0}
\end{eqnarray}
where $\delta$ is the Kronecker delta.
As time $t$ goes, the semi-infinite system deforms by the applied stress.
In the limit of $t \rightarrow \infty$, the deformation eventually stops at a configuration that balances the applied stress and elastic force; namely ${\tilde {\bf v}}(t \rightarrow \infty, \ks) = 0$.
At this stage, the elastic energy stored by the deformation produces a general solution ${\tilde {\bf v}}_{S}(\ks)$ that satisfies,
\begin{eqnarray}
{\tilde {\bf v}}(t\rightarrow \infty, \ks) 
&=& - \delta_{\ks, {\bf 0}} {\bf f}_{s} \int^{\infty}_0 S_{11}(\tau, \ks) d\tau + {\tilde {\bf v}}_{S}(\ks) \nm \\
&=& - \delta_{\ks, {\bf 0}} {\bf f}_{s} S_{11}(z=0, \ks) + {\tilde {\bf v}}_{S}(\ks) \nm \\
&=& 0. \nm
\end{eqnarray}
We obtain
\begin{eqnarray}
  {\tilde {\bf v}}_{S}(\ks) = {\bf f}_{s} S_{11}(z=0, \ks) \delta_{\ks, {\bf 0}} \equiv {\bf v} \delta_{\ks, {\bf 0}}. \label{eq:stress1}
 \end{eqnarray}
Interestingly, this equation indicates that the applied stress is proportional to the constant velocity term.
Note that a general solution from initial conditions of constant velocity, which is 
${\tilde {\bf u}}(0, \ks)=0, {\tilde {\bf v}}(0, \ks) = {\bf v} \delta_{\ks, {\bf 0}}$
is equivalent to Eq. \ref{eq:stress1}.
Namely, the constant shear stress becomes the same as the initial condition in which we start the dynamics by giving the constant velocity to the semi-infinite solid system.

In short, by adding $d{\tilde {\bf u}_g}/dt = {\tilde {\bf v}}_{T} + {\tilde {\bf v}}_{S}$ to the trajectory of the surface layer, we can control the temperature, normal stress and sliding velocity of the semi-infinite system.

\subsubsection{Numerical treatment of thermostat} \label{appendix:thermostat}
We show a numerical recipe to generate random velocity which holds the fluctuation-dissipation theorem in Eq.~\ref{eq:temperature}. An algorithm proposed by Berkowitz\cite{berkowitz1983generalized} is used.
By assuming that ${\tilde v}_{T; \xi}(t, \ks)$ is periodic in an enough long period $P$, the Fourier series expansion yields
\begin{eqnarray}
{\tilde v}_{T; \xi}(t, \ks) = \sum_{n=1}^{\infty} \left(
	a_{\xi, n} \cos(\omega_n t) + b_{\xi, n} \sin(\omega_n t) 
\right),\label{eq:randomT}
\end{eqnarray}
where $\omega = 2\pi n /P$.
The random variables $a_{\xi, n} $ and $b_{\xi, n}$ are assumed to be independent. By inserting Eq.~\ref{eq:randomT} into the left side of Eq. \ref{eq:temperature}, it becomes
\begin{eqnarray}
\langle {\tilde v}_{T; \xi}(t, \ks)   {\tilde v}^*_{T; \xi'}(0,\ks) \rangle 
= &&  \sum_{n=1}^{\infty}
  \langle a_{\xi, n} a^*_{\xi', n}  \rangle \cos(\omega_n t)  \nm \\
&+&\langle b_{\xi, n} a^*_{\xi', n}  \rangle \sin(\omega_n t). \label{eq:lefttemperature}
\end{eqnarray}
We extend the domain $t\ge0$ of $S_{\xi, \xi'}(t, \ks)$ to $\infty \ge t \ge -\infty$ by using $S_{\xi, \xi'}(|t|, \ks)$.
The right side of of Eq. \ref{eq:temperature} is modified as
\begin{eqnarray}
&S_{\xi, \xi'}&(|t|, \ks) \nm \\
&=& \frac{2k_BT}{P} \sum_{n=1}^{\infty} \cos(\omega_n t) \int_{-P/2}^{P/2} S_{\xi, \xi'}(|t'|, \ks)  \cos(\omega_n t') dt' \nm \\
&=& \frac{4k_BT}{P} \sum_{n=1}^{\infty} \cos(\omega_n t) \int_{0}^{P/2} S_{\xi, \xi'}(t', \ks)  \cos(\omega_n t') dt'. \nm 
\end{eqnarray}
Because of the assumption that $P$ is enough large, we can use the cosine transformation
$\mathscr{C}[X](\omega) = 
1/2 \int^{\infty}_{0} (\exp[- \mathrm{i} \omega t] + \exp[ \mathrm{i} \omega t]) X(t) dt$.
Therefore, 
\begin{eqnarray}
S_{\xi, \xi'}(|t|, \ks) = \frac{4k_BT}{P} \sum_{n=1}^{\infty} \mathscr{C}[S_{\xi, \xi'}](\omega_n, \ks) \cos(\omega_n t). \nm \\
\label{eq:righttemperature}
\end{eqnarray}
Inserting Eqs.~\ref{eq:lefttemperature} and \ref{eq:righttemperature} into Eq.~\ref{eq:temperature}, we obtain relations of the random variables required by the fluctuation-dissipation theorem, as
\begin{eqnarray}
  \langle a_{\xi, n} a^*_{\xi', n}  \rangle  &=&  \frac{4k_BT}{P} \mathscr{C}[S_{\xi, \xi'}](\omega_n, \ks) \nm \\
 \langle b_{\xi, n} a^*_{\xi', n}  \rangle &=& 0. \nm
\end{eqnarray}
Because randomness of the ${\tilde v}_{T; \xi}(t, \ks)$, the ensemble average of the magnitudes of $a_{\xi, n}$ and $b_{\xi, n}$ are equivalent: $ \langle a_{\xi, n} a^*_{\xi', n}  \rangle  =  \langle b_{\xi, n} b^*_{\xi', n}  \rangle$.

Then, we construct a covariance matrix $\Sigma$ to generate the random variables $a$ and $b$.
\begin{eqnarray}
\Sigma &=& 
 < 
 \left(
 \begin{array}{c}
 a_{1,n} \\
 b_{1,n} \\
 \vdots \\
 a_{\xi, n} \\
 b_{\xi, n} \\
 \vdots \\
\end{array} 
\right)
\left( a^*_{1,n},  b^*_{1,n}, \cdots, a^*_{\xi,n},  b^*_{\xi,n}, \cdots \right)
 >  \nm \\
 &=&  \frac{4 k_B T}{P} 
 \left(
 \begin{array}{cccccc}
 \mathscr{C}[S_{1, 1}]   &        0                              & \cdots & \mathscr{C}[S_{1, \xi}]  &     0                          & \cdots \\
             0                       & \mathscr{C}[S_{1, 1}]     & \cdots &       0                           & \mathscr{C}[S_{1, \xi}] & \cdots \\
 \vdots                           &  \cdots                             & \cdots & \cdots                         & \cdots                        & \cdots  \\
 \mathscr{C}[S_{\xi, 1}]    &         0                             & \cdots & \mathscr{C}[S_{\xi, \xi}]   &     0                             & \cdots  \\
             0                       & \mathscr{C}[S_{\xi, 1}]       & \cdots &          0                       & \mathscr{C}[S_{\xi, \xi}]   & \cdots   \\
 \vdots         &  \vdots       & \vdots &  \vdots            &  \vdots        &  \vdots \\
\end{array} 
\right) \nm \\
\label{eq:covarianceMat} 
\end{eqnarray}
This matrix is the Hermite matrix, given the fact that $\mathscr{C}[S_{\xi, \xi'}] = \mathscr{C}[S_{\xi', \xi}]^*$.
A stochastic vector ${\bf x} = \left(  A_{1,n},  B_{1,n}, \cdots, A_{\xi,n},  B_{\xi,n}, \cdots  \right)$ that satisfies
Eq.~\ref{eq:covarianceMat} can be generated by considering a multi-variable Gauss distribution $\mathcal{N}$ with the covariance matrix $\Sigma$.
\begin{eqnarray}
\mathcal{N} = \frac{1}{(2\pi)^{D} |\Sigma|^{1/2}} \exp(-\frac{1}{2} {\bf x}^{\dagger} \Sigma^{-1} {\bf x}),
\label{eq:gauss}
\end{eqnarray}
where the number of elements of $\bf{x}$ is $2D$.
We define a matrix $U$ that diagonalizes $\Sigma$ and its eigenvalue matrix $\lambda$.
The variable ${\bf x}^{\dagger} \Sigma^{-1} {\bf x}$ is transformed as,
\begin{eqnarray}
 {\bf x}^{\dagger} \Sigma^{-1} {\bf x} 
 =  {\bf x}^{\dagger} U^{-1} \lambda^{-1} U {\bf x} 
= {\bf y}^{\dagger} \lambda^{-1} {\bf y} = \sum_{i}^{2D}  \frac{|y_i|^2}{\lambda_i},
\label{eq:transform} 
\end{eqnarray}
where ${\bf y} = U {\bf x}$.
Inserting Eq.~\ref{eq:transform} into Eq.~\ref{eq:gauss}, the Gauss distribution becomes,
\begin{eqnarray}
\mathcal{N} = \frac{1}{(2\pi)^{D} |\Sigma|^{1/2}} \prod_{i}^{2D} \exp(-\frac{|y_i|^2}{2 \lambda_i}).
\label{eq:gauss2}
\end{eqnarray}
Equation~\ref{eq:gauss2} indicates that 
$\mathcal{N}$ is expressed by the products of independent Gaussians that have no correlation between the variables.
Therefore, Box-Muller method can be used to generate stochastic variables regulated by the Gauss distribution.
The variable $y_i$ is calculated by
\begin{eqnarray}
y_i = \sqrt{-2 \lambda_i \log \theta_1} \exp(2 \pi \mathrm{i} \theta_2 ), \nm
\end{eqnarray}
where $\theta_{1}$ and $\theta_2$ are uniform random variables ranging from 0.0 to 1.0.
Then, converting ${\bf x} = U^{-1} {\bf y}$, we obtain the thermal velocity terms ${\tilde v}_{T; \xi}(t, \ks)$ via Eq. \ref{eq:randomT}. In this study, we use $P = 2^{19} h$.

%%---------------------------------------------------------------
\subsection{Coupling the QM and GF systems}\label{sec:qmGF MD}
\subsubsection{Add-remove method} \label{subsec:addremove}
\begin{figure}[htbp]
\centering
\includegraphics[width=0.7\linewidth]{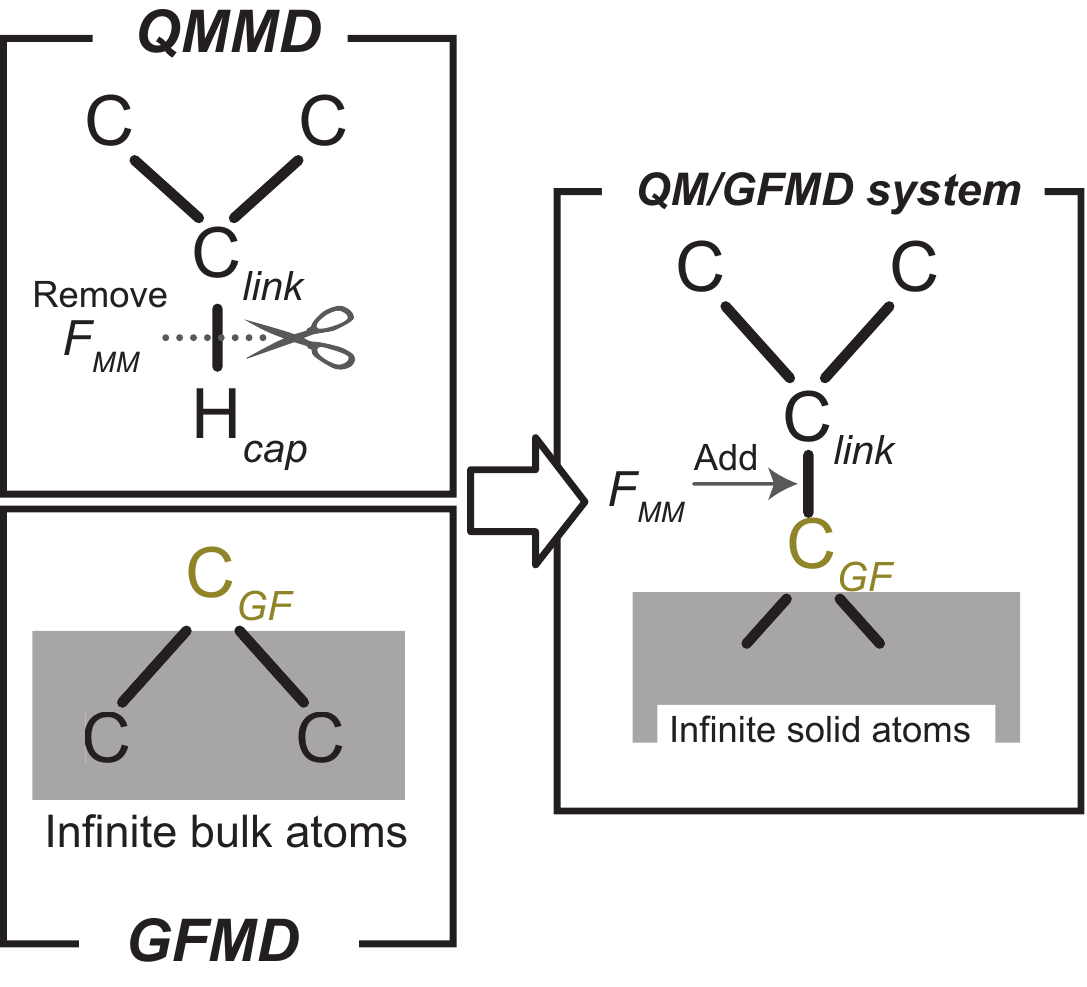} 
\caption{
Schematic representation of the add-remove method in a diamond surface of the QMGF system.
The force $F_{\textrm{MM}}$ indicates a classical force field of the corresponding bond.
}
\label{pic:addremove}
\end{figure}

To couple two systems of different scales, their junction should be bridged smoothly~\cite{venugopalan2017green, otani2006first,ohba2020large, swart2003addremove}.
This study uses an add-remove method, which is one of the hybrid schemes for solids~\cite{swart2003addremove}.
Hydrogens are often used to cap the boundaries of a QM system to stabilize the unsaturated edge atoms, while mechanical contributions such as forces from the artificial cap atoms are eliminated because they should not be present in the junction~\cite{ohba2020large, swart2003addremove}.
Figure~\ref{pic:addremove} shows an outline of the method in the diamond slab, where $\textrm{H}_{\textrm{cap}}$ and  $\textrm{C}_{\textrm{link}}$ denote the cap hydrogens and linked carbons, respectively.
A surface carbon generated by GF MD is indicated by $\textrm{C}_{\textrm{GF}}$.

The add-remove method works as follow:
\begin{enumerate}
\item  $\textrm{H}_{\textrm{cap}}$-$\textrm{C}_{\textrm{link}}$ bonds are removed by subtracting the corresponding classical force fields.
\item  The QM and GF systems are connected with a classical $\textrm{C}_{\textrm{link}}$-$\textrm{C}_{\textrm{GF}}$ bond.
\item  The positions of $\textrm{H}_{\textrm{cap}}$ are located along the projection of the straight line connecting $\textrm{C}_{\textrm{link}}$ and $\textrm{C}_{\textrm{GF}}$. The bond length of $\textrm{H}_{\textrm{cap}}$-$\textrm{C}_{\textrm{link}}$ is fixed at its equilibrium distance.
\item  The forces are corrected due to the constrain of the $\textrm{H}_{\textrm{cap}}$ position.
\end{enumerate}

%---------------------

%---------------------
We describe details of points 3 and 4. 
The constrained position of the cap hydrogen ${\bf r}_{\textrm{cap}}$ are
\begin{eqnarray}
{\bf r}_{\textrm{cap}} = {\bf r}_{\textrm{link}} + r_{{\textrm eq}} \ {\bf u}_{\textrm{link-GF}} \label{eq:constraint} \\
{\bf u}_{\textrm{link-GF}} = \frac{ {\bf r}_{\textrm{GF}} - {\bf r}_{\textrm{link}} }{| {\bf r}_{\textrm{GF}} - {\bf r}_{\textrm{link}} |} \nm
\end{eqnarray}
where ${\bf r}_{\textrm{link}}$ and ${\bf r}_{\textrm{GF}}$  are the positions of $\textrm{C}_{\textrm{link}}$ and $\textrm{C}_{\textrm{GF}}$, respectively.
The symbol $ r_{{\textrm eq}}$ indicates the length of bond
$\textrm{H}_{\textrm{cap}}$-$\textrm{C}_{\textrm{link}}$ fixed at its equilibrium distance.
Due to the ${\bf r}_{\textrm{cap}}$ constrain, the forces should be corrected. We consider a Hamiltonian of the whole system $\mathscr{H}$ including contributions of the add-remove method.
 \begin{eqnarray}
 \mathscr{H} &=& \mathscr{H}_{\textrm{QM}} + \mathscr{H}_{\textrm{GF}} \nm \\
 &&+ \mathscr{H}_{\textrm{add}}({\bf r}_{\textrm{link}}, {\bf r}_{\textrm{GF}}) 
 - \mathscr{H}_{\textrm{remove}}({\bf r}_{\textrm{link}}, {\bf r}_{\textrm{cap}}), \nm
 \end{eqnarray}
 where
$\mathscr{H}_{\textrm{QM}}$ and $\mathscr{H}_{\textrm{GF}}$ indicate the original Hamiltonians of the QM and semi-infinite harmonic oscillator systems as shown in Fig. \ref{pic:fig1}(a), respectively.
The addition and removal operations in Fig.~\ref{pic:addremove} are represented by 
$\mathscr{H}_{\textrm{add}}$ and $\mathscr{H}_{\textrm{remove}}$, which come from the classical interactions of 
$\textrm{C}_{\textrm{link}}$-$\textrm{C}_{\textrm{GF}}$ and $\textrm{H}_{\textrm{cap}}$-$\textrm{C}_{\textrm{link}}$, respectively.
Given the constrained ${\bf r}_{\textrm{cap}}(  {\bf r}_{\textrm{link}}, {\bf r}_{\textrm{GF}} )$ in Eq. \ref{eq:constraint}, the forces acting on $\textrm{C}_{\textrm{GF}}$ and $\textrm{C}_{\textrm{link}}$ atoms are
\begin{eqnarray}
{\bf f}_{\textrm{GF}} &=& - \frac{ \partial ( \ha_{\textrm{GF}} + \ha_{\textrm{add}}) }{\partial {\bf r}_{\textrm{GF}}} \nm \\
                   &&- \frac{ \partial ( \ha_{\textrm{QM}}  - \ha_{\textrm{remove}}) }{\partial {\bf r}_{\textrm{cap}}} 
                       \frac{ \partial {\bf r}_{\textrm{cap}} }{\partial {\bf r}_{\textrm{GF}}}
                       \label{eq:addremove1} \\ 
{\bf f}_{\textrm{link}} &=& - \frac{ \partial ( \ha_{\textrm{QM}} + \ha_{\textrm{add}} - \ha_{\textrm{remove}}) }{\partial {\bf r}_{\textrm{link}}} \nm \\
                  && - \frac{ \partial ( \ha_{\textrm{QM}}  - \ha_{\textrm{remove}}) }{\partial {\bf r}_{\textrm{cap}}} 
                       \frac{ \partial {\bf r}_{\textrm{cap}} }{\partial {\bf r}_{\textrm{link}}} 
                       \label{eq:addremove2} \\ 
%%------
\frac{ \partial {\bf r}_{\textrm{cap}} }{\partial {\bf r}_{\textrm{GF}}}  
                  &=& \frac{ r_{{\textrm eq}}}{| {\bf r}_{\textrm{GF}} - {\bf r}_{\textrm{link}} |} \left(
                  I -  {\bf u}_{\textrm{link-GF}} ( {\bf u}_{\textrm{link-GF}}  )^T 
                  \right)  \nm \\
\frac{ \partial {\bf r}_{\textrm{cap}} }{\partial {\bf r}_{\textrm{link}}}  
                  &=& I - \frac{ \partial {\bf r}_{\textrm{cap}} }{\partial {\bf r}_{\textrm{GF}}}  \nm 
\end{eqnarray}
where $I$ is the unit matrix.
Note that ${\bf u} ({\bf u})^T$ indicates the dyadic product.
The first terms on the right sides of Eqs.~\ref{eq:addremove1} and \ref{eq:addremove2}
consist of interaction forces obtained by GF MD in Eq.~\ref{eq:reducedF} and QM MD simulations along with the add-remove classical force terms, respectively.
The second terms in Eqs.~\ref{eq:addremove1} and \ref{eq:addremove2} come from the constraint of Eq.~\ref{eq:constraint}.
These force terms become zero if the classical force of  $\textrm{H}_{\textrm{cap}}$-$\textrm{C}_{\textrm{link}}$ completely agree with the QM bond.
However, because the classical model cannot reproduce the quantum method perfectly, these constraint-force corrections should be included to keep energy conservation law.

%%---------------------------------------------------------------
\subsubsection{Refresh strategy}
Simulations of sliding friction typically require several hundred thousand steps.
As shown in  Eq.~\ref{eq:reducedF}, the convolution of the GF MD increases its integral time range as time evolves.
This fact induces an accumulation of the integral errors in such a long simulation, leading to inaccurate dynamics and overall instability of the GF MD simulation. 
This subsection provides a remedy for this issue.
%---------------------

\begin{figure}[htbp]
    \centering
    \includegraphics[width=\linewidth]{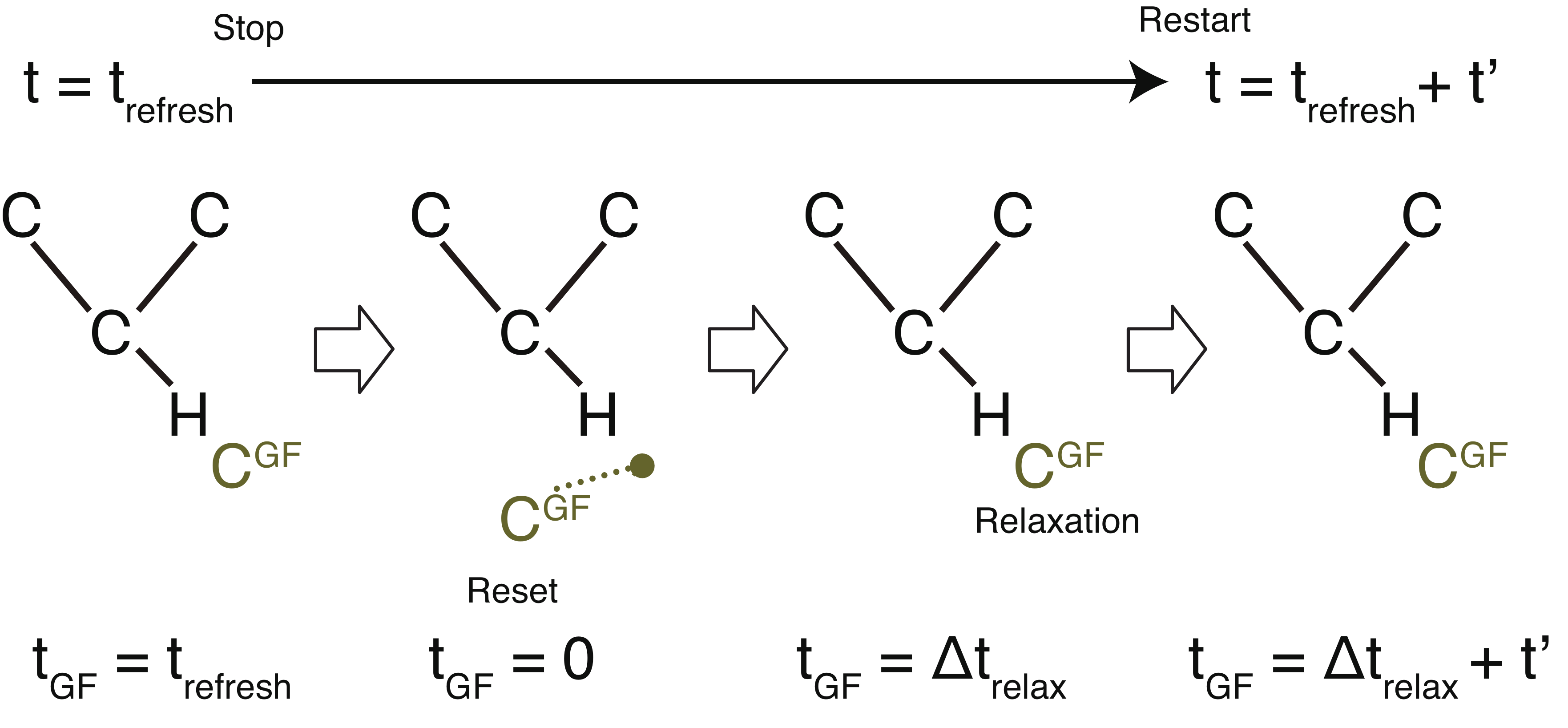}
    \caption{Schematic of the refresh treatment process. 
    The dotted line indicates a spring that is used for the relaxation step presented in the main text.}
    \label{pic:refresh}
\end{figure}

%---------------------

Given that the error comes from the extension of the integral region, an idea would be to reset the convolution before the error cannot be ignored anymore. Figure~\ref{pic:refresh} presents an outline of this treatment, which we call "refresh".
Two clocks $t$ and $t_{\textrm{GF}}$ are prepared for QM and GF MD systems, respectively.
The two clocks advance exactly in the same manner at the start of the QMGF MD simulation. Once they arrive at a user-defined $t_{\textrm{refresh}}$, at which the GF MD numerical error is considered critical, the positions ${\bf r}_{\textrm{GF}}(t_{\textrm{refresh}})$ are saved as anchors in the memory.
At this point, the $t$ clock stops, but only the $t_{\textrm{GF}}$ clock is reset to zero to make the integral range of the convolution zero.
The positions ${\bf r}_{\textrm{GF}}$ are connected to the anchored positions with specific springs to be arranged to their initial positions with respect to ${\bf r}_{\textrm{link}}$.
When we start the $t_{\textrm{GF}}$ clock, but still keep the t clock stopped, the springs pull the $\textrm{C}_{\textrm{GF}}$ in such a way that ${\bf r}_{\textrm{GF}}$ returns to the anchored positions as a result of the relaxation.
After a certain relaxation time $\Delta t_{\textrm{relax}}$, the springs are removed and the $t$ clock starts to run together with $t_{\textrm{GF}}$.

By iterating this refresh every time $t_{\textrm{GF}} = t_{\textrm{refresh}}$, we can perform long and stable GF MD simulations.
This treatment, however, provides artificial effects to $\textrm{C}_{\textrm{link}}$ when the $t$ clock restarts because the velocities of $\textrm{C}_{\textrm{GF}}$ are lost as a consequence of the relaxation.
Nonetheless, because $\textrm{C}_{\textrm{link}}$ are the junction atoms of the hybrid system in the QM bulk region (see Fig.~\ref{pic:fig1}(a)), this error can be regarded as a perturbation that does not to affect surface phenomena if the QM slab model consists of several atomic layers.
 
%%---------------------------------------------------------------
\subsection{Computational details}\label{sec:compdetails}

The internal force matrix  $D$ is calculated by static {\it ab initio} calculations of the diamond bulk, based on density functional theory (DFT) and a DFT linear-response approach to phonons calculation~\cite{qephonon}, performed with the pw.x and ph.x solvers from the Quantum Espresso package~\cite{qe,qe2,qe3}.
The Perdew, Burke, and Ernzerhof generalized gradient approximation is used for the exchange-correlation functional~\cite{pbe}. 
Electronic wave functions are expanded on a plane-wave basis set with a cutoff energy of 25 Ry, and ionic species are described by ultra-soft pseudopotentials~\cite{vanderbilt}. The matrix is approximated so that it only contains elements related to the nearest-neighbor interactions. The off-diagonal elements of the directional indices are also eliminated for the sake of numerical simplicity.

For the add-remove method, the classical force field of the $\textrm{C}_{\textrm{link}}$-$\textrm{C}_{\textrm{GF}}$ bond is set at the value of the corresponding element of the internal force matrix.
The $\textrm{C}_{\textrm{link}}$-$\textrm{H}_{\textrm{cap}}$ spring constant is estimated from {\it ab initio} static calculations performed on a fully H-terminated 2$\times$1 (111) diamond slab of 12 atomic layers.
%Change of the force is observed when the H position shifts on the fixed diamond surface.
The estimated spring constant of the surface normal direction is $0.2575$~Ht/bohr, while for the surface lateral direction is $0.0365$~Ht/bohr. The stable bond length of $\textrm{C}_{\textrm{link}}$-$\textrm{H}_{\textrm{cap}}$ is $r_{{\textrm eq}} = 2.1043$~bohr.
%The Green's function was derived from Eqs.~\ref{eq:effectiveK} and \ref{eq:gmain} via the damped Newton algorithm.

We implemented the QMGF MD hybrid method into the Car-Parrinello solver cp.x. Time development of cp.x is solved by the Verlet method, which does not use velocities of atoms explicitly.
On the other hand, the GF MD uses the general solutions to impose the temperature and stress by adding the velocity corrections ${\bf v}_T$ and ${\bf v}_S$. In order to merge the velocity correction into the QM MD algorithm, we used the leap-frog method that explicitly leverages the velocity term but is compatible to the Verlet method, as follows.
\begin{eqnarray}
%{\bf p}_{\textrm{GF}}(t) 
%&\leftarrow& {\bf p}_{\textrm{GF}} (t-\frac{1}{2}h)  + \frac{1}{2} h {\bf f}_{\textrm{GF}} (t-h) 
%\nm \\
%{\bf v}(t) 
%&\leftarrow& M^{-1} {\bf p}_{\textrm{GF}} (t)
%\nm \\
{\bf p}_{\textrm{GF}}(t+\frac{h}{2}) 
&\leftarrow&  {\bf p}_{\textrm{GF}} (t-\frac{h}{2})  +  h  {\bf f}_{\textrm{GF}} (t) 
\nm \\
{\bf v}_{\textrm{GF}}(t+\frac{h}{2}) 
&\leftarrow&  M^{-1} {\bf p}_{\textrm{GF}}(t+\frac{h}{2})  +   {\bf v}_{T} (t) + {\bf v}_{S}
\nm \\
{\bf r}_{\textrm{GF}}(t+h) 
&\leftarrow&  {\bf r}_{\textrm{GF}}(t)  + h {\bf v}_{\textrm{GF}}(t+\frac{h}{2}),
\nm 
\end{eqnarray}
where ${\bf p}_{\textrm{GF}}, {\bf v}_{\textrm{GF}}$, and ${\bf r}_{\textrm{GF}}$ are momentum, velocity, and position vectors of the GF MD atoms respectively.
The time step is set to $h=0.1$~fs, and the reduced force ${\bf f}_{\textrm{GF}}$ is calculated by Eq.~\ref{eq:reducedF}. %\red{$h$ is set to ** fs.}
The parameters of mTILT are $B=11$ and $N=60$, which is equivalent to the 121 integral points in the contour.
The singular points of the Green's function are searched by evaluating its first and second derivatives on the imaginary axis. 
 %Figure * shows an example of the singular points of the Green's function.
In the refresh treatment, we use $t_{\textrm{refresh}} = 50,000 h$, the anchor spring constant is 0.05~Ht/bohr for all the x, y, and z directions, and the relax time is $2,000 h$. Temperature is set to 300 K by the thermostat of the GF MD method.

The QM ions are thermalized by applying a Nosé-Hoover thermostat with a frequency of 80 Thz and imposing an average electronic kinetic energy of 0.25 atomic units on the electron degrees of freedom. The electronic mass and the time step of the molecular dynamics are selected to be 100 and 4 atomic units, respectively.
%The atomic temperatures are reported in Fig.~1 of Supplementary Information.
At the beginning of our dynamic simulations, the CP solver is employed to obtain the ground state energy of the electronic wave functions with the steepest descent algorithm. Subsequently, the hybrid QMGF MD code is used to carry out the dynamic simulation. 
The computational parameters adopted for the CP scheme have been carefully selected to achieve good accordance between the temperatures of the QM and GF atoms during the dynamics for the system under study.

\subsection{Summary of the QMGF method}\label{sec:compdetails}

%\subsection*{QM GF molecular dynamics in a nutshell}\label{sec:nutshell}

\begin{figure*}[htbp]
\centering
\includegraphics[width=\linewidth]{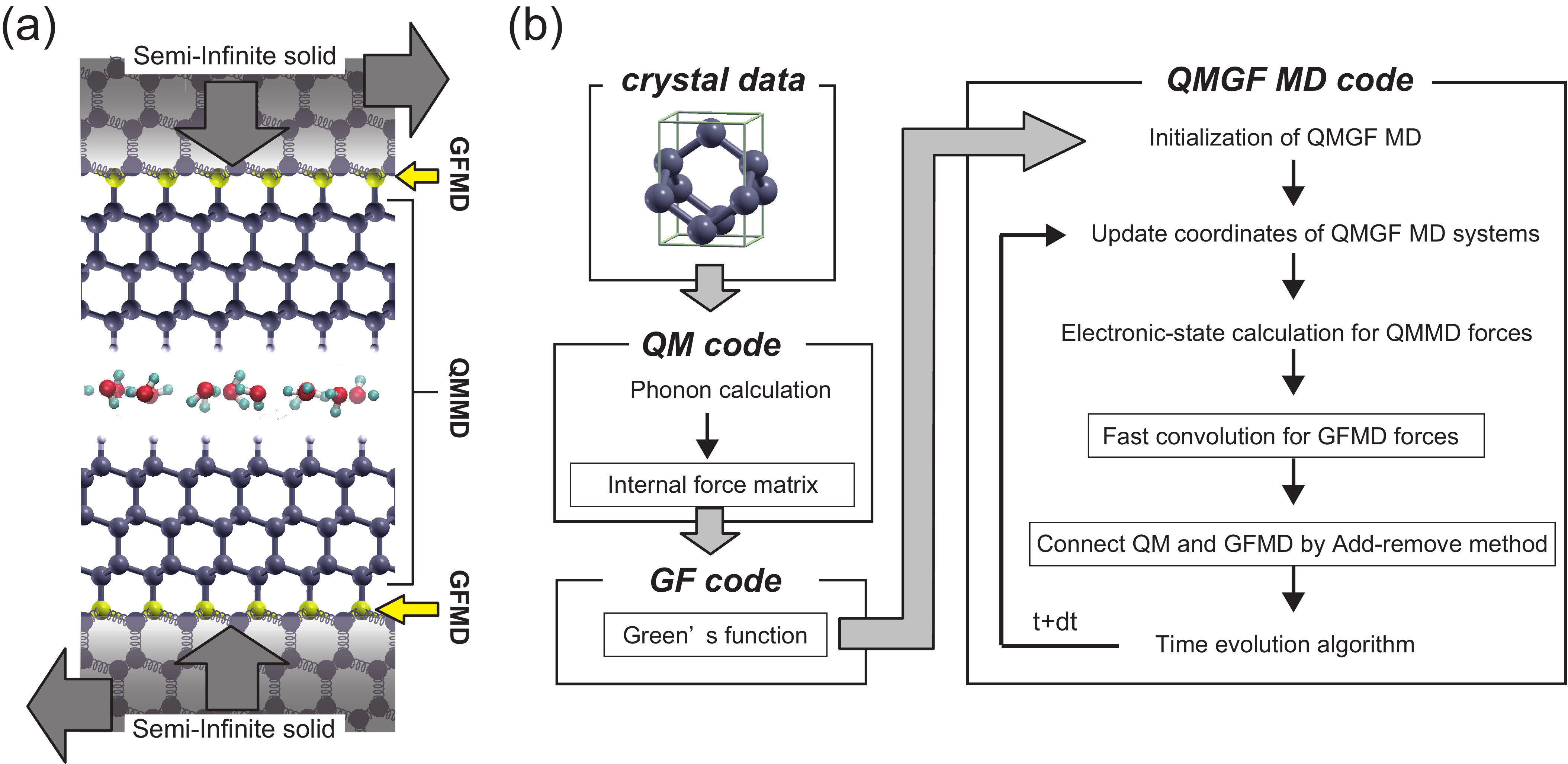} 
\caption{
Representation of a frictional interface described by the QMGF MD hybrid scheme. The GF MD atoms at the boundary of the QM region are colored in yellow, the semi-infinite bulks are represented as coupled harmonic oscillators (a). The workflow of the developed QMGF MD program by linking an open-source {\emph ab initio} code\cite{qe} to the in-house developed GF MD code (b).
}
\label{pic:fig1}
\end{figure*}

A pictorial representation of the hybrid QMGF MD scheme and its application to a prototypical tribochemistry system is offered in Figure~\ref{pic:fig1}. The chemically active part of the system consists of two surfaces in contact and some molecules eventually confined between them (Fig.~\ref{pic:fig1}(a)). The inclusion of the electronic degrees of freedom is necessary to capture quantum effects, such as the Pauli repulsion at the short distances imposed by the applied load and the enhanced chemical reactivity of confined species, which deeply affect the tribological behavior. The two semi-infinite bulks are described by a collection of an infinite number of harmonic oscillators of first-principles derived spring constants. 
Their effect is fully taken into account by the surface atoms indicated in yellow. 
The basic idea of GF MD is, in fact, that all the internal modes of an elastic solid can be integrated out and substituted by effective interactions~\cite{kubo-1966, campana2006practical}. In this way, only the trajectories of the quantum atoms and the surface atoms treated by the GF MD are needed, and no other bulk atoms are needed to be included in the simulation. \newline
The workflow of the QMGF MD method is shown in Fig.~\ref{pic:fig1}(b).
The model for the bulk crystal is constructed, and static first-principles calculations are used to obtain the force matrix, which is used to calculate the Green's function. The QM and GF systems are finally coupled via an add-remove scheme~\cite{swart2003addremove}. 

\section{Results for diamond interfaces}\label{sec:results}

\begin{figure*}
    \centering
    \includegraphics[width=\linewidth]{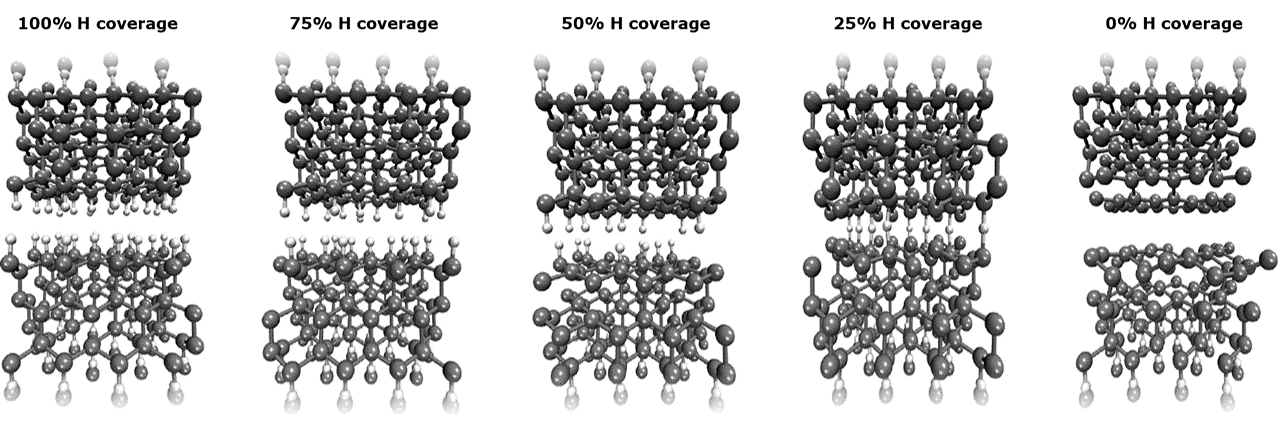}
    \caption{Lateral view of the diamond on diamond systems after 10 ps of sliding motion. The surface energy increases with the number of unsaturated carbon atoms, producing higher adhesion and smaller separation. When hydrogen is completely removed, a partial graphitization of the interface can be recognized, and the interfacial separation becomes similar to the interlayer distance of graphite (3.3 \AA).}
    \label{fig:main-sys}
\end{figure*}

We employed our QMGF MD solver to study the sliding interface between two diamond crystals and quantitatively estimate the friction coefficient considering different concentrations of H atoms on the two mated surfaces. We focused our attention on the C(111) surface, the most accessible cleavage plane of diamond, and modeled the diamond-diamond interface by adopting a supercell with (4$\times$2) in-plane size, containing two-faced slabs, each constituted of three bilayers of carbon atoms.
The slabs are externally passivated by hydrogen atoms and the GF atoms are linked to these capping atoms, as described in the method section.
The interfacial region, where the two surfaces are faced, contains hydrogen atoms in different concentrations and randomly distributed. In Fig.~\ref{fig:main-sys} a lateral view of all the considered systems after 10 ps of sliding is reported.

We performed molecular dynamics simulations at a temperature of 300~K with an external load of 5~GPa for a time interval of $\sim$ 50~ps.
To generate the sliding motion, we applied shear stresses of 1~GPa along the $x$~direction by applying external lateral forces in opposite directions on the GF MD atoms of each slab.
As described in Sec. \ref{subsec:thermo_baro},
the surface slabs slide against each other at constant velocity if there is no friction force, 
because  the condition of the constant shear stress is equivalent to a situation where
we start the friction test by imposing a relative velocity
on the semi-infinite solids.

\subsection*{Effects of interfacial adhesion on kinetic friction}

Three values of the H-coverage, $\theta$, turned out to be high enough to enable the sliding motion under the effects of the applied lateral forces. Instead, the other coverages were too low to prevent chemical bonds from forming across the interface, which impeded the lateral displacement. 

\begin{table}[htbp]
\begin{center}
\begin{tabular}{|M{1.0cm}||M{1.0cm}|M{1.0cm}|M{1.6cm}|M{1.0cm}|}
    \hline
    \multicolumn{5}{|c|}{Quantities derived from the sliding dynamics} \\
    \hline
      \thead{$\theta$ }
    & \thead{$\langle d_{C}\rangle$ }
    & \thead{$\langle v_{x}\rangle$ }
    & \thead{$\langle F_{x}^{k}/A\rangle$ }
    & \thead{$\mu_k$}\\
    \hline
    100\% & 4.04 & 51 & 0.15  & 0.03\\
    75\% & 3.72 & 51 & 0.21  & 0.05\\
    50\% & 3.25 & 48 & 0.27  & 0.06\\
\hline
\end{tabular}
\end{center}
    \caption{Results of the QMGF MD simulations. For each hydrogen coverage, $\theta$, the averages of the surface separation $\langle d_{C} \rangle$, sliding velocity $\langle v_{x}\rangle$, kinetic friction force per unit area $\langle F_{x}^{k}/A\rangle$, and the kinetic friction coefficient $\mu_k$ are calculated. $\langle F_{x}^{k}\rangle$ is calculated as the time average of the interfacial forces acting on the GF atoms along the $x$ direction. These are the forces appearing in the convolution integral in Eq.\ref{eq:reducedF}. The kinetic friction coefficient is calculated as the ratio between $\langle F_{x}^{k}\rangle$ and the applied load on the GF atoms.}
    \label{tab:mu_table}
\end{table}

\begin{figure}[htbp]
    \centering
    \includegraphics[width=\linewidth]{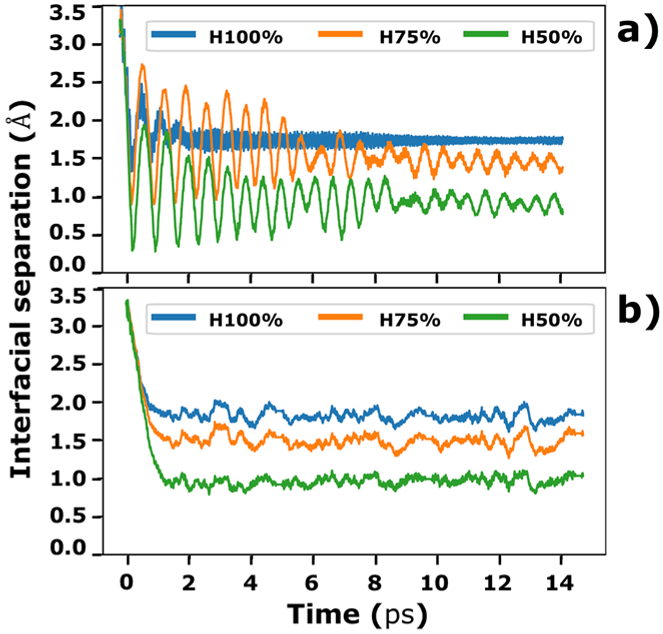}
    \caption{Interfacial separation for the 100\%, 75\% and 50\% passivated systems during the QM MD (a) and QMGF MD (b) simulation of sliding. The surface separation is calculated by considering the $z$ coordinates of the hydrogen atoms.}
    \label{fig:delta_z}
\end{figure}

The results in tab. \ref{tab:mu_table} highlight the effect of surface passivation on kinetic friction. A decrease in hydrogen coverage always results in a friction increase with a corresponding reduction in sliding velocity and average slab separation, also shown in Fig~\ref{fig:delta_z}(a).
This behavior can be explained in terms of the chemical reactivity of the facing diamond surfaces. When H atoms are removed from the diamond surface, the terminal C atoms expose dangling bonds, which are very reactive. The dangling bonds of two surfaces in contact interact and cause a significant increase in the adhesive friction of the system. \newline
The calculated friction coefficients are in agreement with diamond-on-diamond experiments in an air environment, where $\mu_k$ ranges between 0.01 and 0.1~ \cite{Tabor-1979, Hayward-1991, Erdermir-2018}.
This extremely low friction has been detected for different surfaces of diamonds, e.g., the (100)~\cite{Samuels-1988, Hayward-1991, Feng-1992, Germann-1993}, (110)~\cite{Samuels-1988} and also for nanocrystalline diamond films and diamond-like carbon (DLC) employed as coatings in technological applications~\cite{Erdermir-2018, Ajikumar-2019}.
In particular, for the (111) diamond face, experimental results predict a friction coefficient of the order of 0.05~\cite{Germann-1993}, which is almost constant along any possible sliding direction and independent from the applied load. These experimental results are in agreement with the values extracted from our simulations. \newline

The critical role of surface passivation by hydrogen or by environmental molecules, such as water molecules, for achieving low friction coefficients has been highlighted by different experimental works, both for diamond and DLC films~\cite{Konicek-2008, Wang-2013, Cui-2014}. Static first-principles calculations have quantified this effect on the ideal interfacial shear strength~\cite{Zilibotti-2009, Zilibotti-2011, Bouchet-2012, Kuwahara-2017} and {\it ab initio} MD simulations allowed us to monitor the tribochemical processes that lead to the diamond surface passivation by water during sliding~\cite{Zilibotti-2013}. As a further step, we are now able to assess kinetic friction coefficients using QMGF MD simulations thanks to the capability to provide proper control of temperature, mechanical stresses, and energy dissipation in non-equilibrium conditions.\newline

We further reduce the H coverage by considering a passivation of 25\% and an H-free interface. In the former case, the sliding motion occurs only in the first stages of the simulations but then the slabs interlock due to the formation of chemical bonds across the interface, which are not broken by the applied lateral force. \newline
On the contrary, in the clean interface, the motion occurs with no interlocking. In Fig.~\ref{fig:main-sys}, a snapshot of the system acquired during the simulation reveals that a graphitization of the surfaces is taking place due to a partial re-hybridization of the carbon surface bonds from $sp^3$ to $sp^2$. This determines the formation of interfacial graphene layers, which become almost detached from the diamond slabs. %subsequent snapshots \newline
 
 The graphitization mechanism of carbon films induced by sliding has been observed experimentally\cite{Liu-1996, Voevodin-1996, DeBarrosBouchet-2015}, and by MD simulations\cite{Pastewka2010, Kuwahara-2017}, and it was related to the ultralow friction coefficient of diamond. The realistic simulations here performed clarify that the condition to achieve an ultralow friction coefficient is to make the surface-surface interaction change from chemical to physical. This condition can be realized either by increasing the level of passivation above a limiting value where the Pauli repulsion makes the surface separation high enough to inhibit the formation of bonds across the interface or by decreasing the passivation below a threshold value, where the surface graphitization takes place. 
 
To evaluate the effects of the elastic properties of the semi-infinite bulks on the sliding dynamics and friction coefficients, we compare the results of QMGF MD with those of QM MD, obtained by decoupling the atomistic slabs from the bath of harmonic oscillators. Figure~\ref{fig:delta_z} shows the vertical separations of the diamond surfaces during the QM MD (a) and QMGF MD (b) simulations, where the same initial relative position of the two surfaces, external load, and shear are considered. In the absence of a proper description of the inertia and elasticity of the semi-infinite bulks through the GF MD, we observe a marked bumping of the surfaces. The bumping oscillations fade away quickly in the case of complete superficial passivation while they persist for the 75\% and 50\% cases. While the average values of the surface separations are similar for the two kinds of simulations, the lack of contact between the surfaces after each bouncing event produces large system accelerations in the QM MD simulations and makes any quantitative estimate of frictional parameters absolutely not meaningful.

%%-----------------------------------------------------------------------------------------------------%%
\section{Conclusions} \label{sec:conclusion}

Classical tribology was developed in the context of mechanical engineering, where friction forces are predicted on the basis of analytical models of contact mechanics. With the advent of nanotribology, it became possible to probe the tribological behavior of a single nano-asperity and, thanks to the increased power of supercomputers, reproduce it with fully atomistic models. It was thus shown that the atomic-scale surface roughness leads to dramatic deviations from continuum theory.\cite{Robbins-2005} Then the need to go beyond the atomistic description and consider also the electrons at the nano-asperity contacts emerged in the context of tribochemistry. Tight-binding MD and then more accurate, but computationally expensive, {\it ab initio}  MD  were introduced in the field of computational tribology to overcome  the limited reliability of force fields in describing stress-assisted reactions (a comprehensive review on computational tribochemistry can be found here \cite{Huong-2021}).

All the above-described approaches suffer from the limitation of using slabs of finite thickness, thus the energy introduced in the system through the application of external forces is "artificially" removed by thermostats that mimic the effects of the thermal bath consisting in the real systems of the infinite degrees of freedom of the bulk. This approximation makes any estimation of energy dissipation nonsense and prevents a full understanding of the interplay of adhesive and phononic contribution to fiction. To overcome these limitations, we developed a multiscale method, the QMGF method, that links {\it ab initio} to Green's function MD. \\
We applied it to calculate the kinetic friction coefficient of two semi-infinite diamond bulks in contact, obtaining results in close agreement with experiments. We found that the friction coefficient, friction mechanisms and interface morphology strongly depend on the degree of passivation of the diamond surfaces. We observe a superlubric regime at high H coverages, while below a threshold coverage, covalent bonds are established across the interface that causes the surface interlocking. This regime persists until the concentration of adsorbates becomes low enough to allow for a shear-induced change of hybridization of the surface carbon atoms from sp3 to sp2. Thanks to surface graphitization, the sliding motion is recovered. Our results indicate that this phenomenon can occur only when passivating species are almost absent from the interface; indeed, we observed graphitization for a clean diamond interface. \newline
The above results, point at the great potentiality of the QMGF method to provide highly accurate insights into interface phenomena in non-equilibrium conditions.
This method may open the way to the investigation of other multiscale phenomena, where the infinite number of the bulk degrees of freedom, usually neglected in {\it ab initio} MD, is key in determining the system response to an external stimulus.

%%-----------------------------------------------------------------------------------------------------%%

%--------------------------------------------------------------------
\begin{acknowledgments}
These results are part of the ”Advancing Solid Interface and Lubricants by First Principles Material Design (SLIDE)” project that has received funding from the European Research Council (ERC) under the European Union’s Horizon 2020 research and innovation program (Grant agreement No. 865633). 
We thank Dr. C. Cavazzoni for the help in implementing the QMGF MD method within the cp.x code.
The pictures in the present paper are created with the help of XCrySDen~\cite{KOKALJ-1999, KOKALJ-2003}, and Matplotlib~\cite{Hunter-2007}. 
\end{acknowledgments}

%\section*{Ethics declarations}
%The Authors declare no Competing Financial or Non-Financial Interests

%\section*{Data Availability}
%All data were available from the corresponding authors upon reasonable request.

%\section*{Code Availability}
%The related codes are available from the corresponding authors upon reasonable request.

\section*{Author Contributions}
The research was conceived by MCR and SK, supervision and project administration by MCR. The Green's Function subroutine was developed by SK and NK. Its linking with AIMD was implemented by MCR's group and SK. 
The simulations were carried out by AP and GL. All authors discussed the results and contributed to writing the manuscript.

%--------------------------------------------------------------------

%\begin{thebibliography}{30}

\bibliographystyle{apsrev4-1}
\bibliography{biblio_main.bib}

\begin{thebibliography}{67}%
\makeatletter
\providecommand \@ifxundefined [1]{%
 \@ifx{#1\undefined}
}%
\providecommand \@ifnum [1]{%
 \ifnum #1\expandafter \@firstoftwo
 \else \expandafter \@secondoftwo
 \fi
}%
\providecommand \@ifx [1]{%
 \ifx #1\expandafter \@firstoftwo
 \else \expandafter \@secondoftwo
 \fi
}%
\providecommand \natexlab [1]{#1}%
\providecommand \enquote  [1]{``#1''}%
\providecommand \bibnamefont  [1]{#1}%
\providecommand \bibfnamefont [1]{#1}%
\providecommand \citenamefont [1]{#1}%
\providecommand \href@noop [0]{\@secondoftwo}%
\providecommand \href [0]{\begingroup \@sanitize@url \@href}%
\providecommand \@href[1]{\@@startlink{#1}\@@href}%
\providecommand \@@href[1]{\endgroup#1\@@endlink}%
\providecommand \@sanitize@url [0]{\catcode `\\12\catcode `\$12\catcode
  `\&12\catcode `\#12\catcode `\^12\catcode `\_12\catcode `\%12\relax}%
\providecommand \@@startlink[1]{}%
\providecommand \@@endlink[0]{}%
\providecommand \url  [0]{\begingroup\@sanitize@url \@url }%
\providecommand \@url [1]{\endgroup\@href {#1}{\urlprefix }}%
\providecommand \urlprefix  [0]{URL }%
\providecommand \Eprint [0]{\href }%
\providecommand \doibase [0]{http://dx.doi.org/}%
\providecommand \selectlanguage [0]{\@gobble}%
\providecommand \bibinfo  [0]{\@secondoftwo}%
\providecommand \bibfield  [0]{\@secondoftwo}%
\providecommand \translation [1]{[#1]}%
\providecommand \BibitemOpen [0]{}%
\providecommand \bibitemStop [0]{}%
\providecommand \bibitemNoStop [0]{.\EOS\space}%
\providecommand \EOS [0]{\spacefactor3000\relax}%
\providecommand \BibitemShut  [1]{\csname bibitem#1\endcsname}%
\let\auto@bib@innerbib\@empty
%</preamble>
\bibitem [{\citenamefont {Holmberg}\ and\ \citenamefont
  {Erdemir}(2017)}]{Holmberg-2017}%
  \BibitemOpen
  \bibfield  {author} {\bibinfo {author} {\bibfnamefont {K.}~\bibnamefont
  {Holmberg}}\ and\ \bibinfo {author} {\bibfnamefont {A.}~\bibnamefont
  {Erdemir}},\ }\href {\doibase 10.1007/s40544-017-0183-5} {\bibfield
  {journal} {\bibinfo  {journal} {Friction}\ }\textbf {\bibinfo {volume} {5}},\
  \bibinfo {pages} {263} (\bibinfo {year} {2017})}\BibitemShut {NoStop}%
\bibitem [{\citenamefont {Wolloch}\ \emph {et~al.}(2018)\citenamefont
  {Wolloch}, \citenamefont {Levita}, \citenamefont {Restuccia},\ and\
  \citenamefont {Righi}}]{Wolloch-2018}%
  \BibitemOpen
  \bibfield  {author} {\bibinfo {author} {\bibfnamefont {M.}~\bibnamefont
  {Wolloch}}, \bibinfo {author} {\bibfnamefont {G.}~\bibnamefont {Levita}},
  \bibinfo {author} {\bibfnamefont {P.}~\bibnamefont {Restuccia}}, \ and\
  \bibinfo {author} {\bibfnamefont {M.~C.}\ \bibnamefont {Righi}},\ }\href
  {\doibase 10.1103/PhysRevLett.121.026804} {\bibfield  {journal} {\bibinfo
  {journal} {Phys. Rev. Lett.}\ }\textbf {\bibinfo {volume} {121}},\ \bibinfo
  {pages} {026804} (\bibinfo {year} {2018})}\BibitemShut {NoStop}%
\bibitem [{\citenamefont {Hsu}\ \emph {et~al.}(2002)\citenamefont {Hsu},
  \citenamefont {Zhang},\ and\ \citenamefont {Yin}}]{Hsu-2002}%
  \BibitemOpen
  \bibfield  {author} {\bibinfo {author} {\bibfnamefont {S.~M.}\ \bibnamefont
  {Hsu}}, \bibinfo {author} {\bibfnamefont {J.}~\bibnamefont {Zhang}}, \ and\
  \bibinfo {author} {\bibfnamefont {Z.}~\bibnamefont {Yin}},\ }\href {\doibase
  10.1023/A:1020112901674} {\bibfield  {journal} {\bibinfo  {journal}
  {Tribology Letters}\ }\textbf {\bibinfo {volume} {13}},\ \bibinfo {pages}
  {131} (\bibinfo {year} {2002})}\BibitemShut {NoStop}%
\bibitem [{\citenamefont {Zilibotti}\ \emph {et~al.}(2013)\citenamefont
  {Zilibotti}, \citenamefont {Corni},\ and\ \citenamefont
  {Righi}}]{Zilibotti-2013}%
  \BibitemOpen
  \bibfield  {author} {\bibinfo {author} {\bibfnamefont {G.}~\bibnamefont
  {Zilibotti}}, \bibinfo {author} {\bibfnamefont {S.}~\bibnamefont {Corni}}, \
  and\ \bibinfo {author} {\bibfnamefont {M.~C.}\ \bibnamefont {Righi}},\ }\href
  {\doibase 10.1103/PhysRevLett.111.146101} {\bibfield  {journal} {\bibinfo
  {journal} {Phys. Rev. Lett.}\ }\textbf {\bibinfo {volume} {111}},\ \bibinfo
  {pages} {146101} (\bibinfo {year} {2013})}\BibitemShut {NoStop}%
\bibitem [{\citenamefont {James}\ \emph {et~al.}(2012)\citenamefont {James},
  \citenamefont {Adams}, \citenamefont {Bolm}, \citenamefont {Braga},
  \citenamefont {Collier}, \citenamefont {Friščić}, \citenamefont
  {Grepioni}, \citenamefont {Harris}, \citenamefont {Hyett}, \citenamefont
  {Jones}, \citenamefont {Krebs}, \citenamefont {Mack}, \citenamefont {Maini},
  \citenamefont {Orpen}, \citenamefont {Parkin}, \citenamefont {Shearouse},
  \citenamefont {Steed},\ and\ \citenamefont {Waddell}}]{Stuart-2012}%
  \BibitemOpen
  \bibfield  {author} {\bibinfo {author} {\bibfnamefont {S.~L.}\ \bibnamefont
  {James}}, \bibinfo {author} {\bibfnamefont {C.~J.}\ \bibnamefont {Adams}},
  \bibinfo {author} {\bibfnamefont {C.}~\bibnamefont {Bolm}}, \bibinfo {author}
  {\bibfnamefont {D.}~\bibnamefont {Braga}}, \bibinfo {author} {\bibfnamefont
  {P.}~\bibnamefont {Collier}}, \bibinfo {author} {\bibfnamefont
  {T.}~\bibnamefont {Friščić}}, \bibinfo {author} {\bibfnamefont
  {F.}~\bibnamefont {Grepioni}}, \bibinfo {author} {\bibfnamefont {K.~D.~M.}\
  \bibnamefont {Harris}}, \bibinfo {author} {\bibfnamefont {G.}~\bibnamefont
  {Hyett}}, \bibinfo {author} {\bibfnamefont {W.}~\bibnamefont {Jones}},
  \bibinfo {author} {\bibfnamefont {A.}~\bibnamefont {Krebs}}, \bibinfo
  {author} {\bibfnamefont {J.}~\bibnamefont {Mack}}, \bibinfo {author}
  {\bibfnamefont {L.}~\bibnamefont {Maini}}, \bibinfo {author} {\bibfnamefont
  {A.~G.}\ \bibnamefont {Orpen}}, \bibinfo {author} {\bibfnamefont {I.~P.}\
  \bibnamefont {Parkin}}, \bibinfo {author} {\bibfnamefont {W.~C.}\
  \bibnamefont {Shearouse}}, \bibinfo {author} {\bibfnamefont {J.~W.}\
  \bibnamefont {Steed}}, \ and\ \bibinfo {author} {\bibfnamefont {D.~C.}\
  \bibnamefont {Waddell}},\ }\href {\doibase 10.1039/C1CS15171A} {\bibfield
  {journal} {\bibinfo  {journal} {Chem. Soc. Rev.}\ }\textbf {\bibinfo {volume}
  {41}},\ \bibinfo {pages} {413} (\bibinfo {year} {2012})}\BibitemShut
  {NoStop}%
\bibitem [{\citenamefont {Fri{\v{s}}{\v{c}}i{\'{c}}}\ \emph
  {et~al.}(2013)\citenamefont {Fri{\v{s}}{\v{c}}i{\'{c}}}, \citenamefont
  {Halasz}, \citenamefont {Beldon}, \citenamefont {Belenguer}, \citenamefont
  {Adams}, \citenamefont {Kimber}, \citenamefont {Honkim{\"a}ki},\ and\
  \citenamefont {Dinnebier}}]{Friscic-2013}%
  \BibitemOpen
  \bibfield  {author} {\bibinfo {author} {\bibfnamefont {T.}~\bibnamefont
  {Fri{\v{s}}{\v{c}}i{\'{c}}}}, \bibinfo {author} {\bibfnamefont
  {I.}~\bibnamefont {Halasz}}, \bibinfo {author} {\bibfnamefont {P.~J.}\
  \bibnamefont {Beldon}}, \bibinfo {author} {\bibfnamefont {A.~M.}\
  \bibnamefont {Belenguer}}, \bibinfo {author} {\bibfnamefont {F.}~\bibnamefont
  {Adams}}, \bibinfo {author} {\bibfnamefont {S.~A.}\ \bibnamefont {Kimber}},
  \bibinfo {author} {\bibfnamefont {V.}~\bibnamefont {Honkim{\"a}ki}}, \ and\
  \bibinfo {author} {\bibfnamefont {R.~E.}\ \bibnamefont {Dinnebier}},\ }\href
  {\doibase 10.1038/nchem.1505} {\bibfield  {journal} {\bibinfo  {journal}
  {Nature Chemistry}\ }\textbf {\bibinfo {volume} {5}},\ \bibinfo {pages} {66}
  (\bibinfo {year} {2013})}\BibitemShut {NoStop}%
\bibitem [{\citenamefont {Ta}\ \emph {et~al.}(2021{\natexlab{a}})\citenamefont
  {Ta}, \citenamefont {Tran}, \citenamefont {Tieu}, \citenamefont {Zhu},
  \citenamefont {Yu},\ and\ \citenamefont {Ta}}]{nam-van-tran-2021}%
  \BibitemOpen
  \bibfield  {author} {\bibinfo {author} {\bibfnamefont {H.~T.~T.}\
  \bibnamefont {Ta}}, \bibinfo {author} {\bibfnamefont {N.~V.}\ \bibnamefont
  {Tran}}, \bibinfo {author} {\bibfnamefont {A.~K.}\ \bibnamefont {Tieu}},
  \bibinfo {author} {\bibfnamefont {H.}~\bibnamefont {Zhu}}, \bibinfo {author}
  {\bibfnamefont {H.}~\bibnamefont {Yu}}, \ and\ \bibinfo {author}
  {\bibfnamefont {T.~D.}\ \bibnamefont {Ta}},\ }\href {\doibase
  10.1021/acs.jpcc.1c03725} {\bibfield  {journal} {\bibinfo  {journal} {The
  Journal of Physical Chemistry C}\ }\textbf {\bibinfo {volume} {125}},\
  \bibinfo {pages} {16875} (\bibinfo {year} {2021}{\natexlab{a}})}\BibitemShut
  {NoStop}%
\bibitem [{\citenamefont {Kajita}\ and\ \citenamefont
  {Righi}(2016)}]{Kajita-2016}%
  \BibitemOpen
  \bibfield  {author} {\bibinfo {author} {\bibfnamefont {S.}~\bibnamefont
  {Kajita}}\ and\ \bibinfo {author} {\bibfnamefont {M.}~\bibnamefont {Righi}},\
  }\href {http://www.sciencedirect.com/science/article/pii/S0008622316301713}
  {\bibfield  {journal} {\bibinfo  {journal} {Carbon}\ }\textbf {\bibinfo
  {volume} {103}},\ \bibinfo {pages} {193} (\bibinfo {year}
  {2016})}\BibitemShut {NoStop}%
\bibitem [{\citenamefont {Kuwahara}\ \emph
  {et~al.}(2017{\natexlab{a}})\citenamefont {Kuwahara}, \citenamefont {Moras},\
  and\ \citenamefont {Moseler}}]{Moseler-2017}%
  \BibitemOpen
  \bibfield  {author} {\bibinfo {author} {\bibfnamefont {T.}~\bibnamefont
  {Kuwahara}}, \bibinfo {author} {\bibfnamefont {G.}~\bibnamefont {Moras}}, \
  and\ \bibinfo {author} {\bibfnamefont {M.}~\bibnamefont {Moseler}},\ }\href
  {\doibase 10.1103/PhysRevLett.119.096101} {\bibfield  {journal} {\bibinfo
  {journal} {Phys. Rev. Lett.}\ }\textbf {\bibinfo {volume} {119}},\ \bibinfo
  {pages} {096101} (\bibinfo {year} {2017}{\natexlab{a}})}\BibitemShut
  {NoStop}%
\bibitem [{\citenamefont {Restuccia}\ and\ \citenamefont
  {Righi}(2016)}]{Restuccia-2016}%
  \BibitemOpen
  \bibfield  {author} {\bibinfo {author} {\bibfnamefont {P.}~\bibnamefont
  {Restuccia}}\ and\ \bibinfo {author} {\bibfnamefont {M.}~\bibnamefont
  {Righi}},\ }\href {\doibase 10.1016/j.carbon.2016.05.025} {\bibfield
  {journal} {\bibinfo  {journal} {Carbon}\ }\textbf {\bibinfo {volume} {106}},\
  \bibinfo {pages} {118 } (\bibinfo {year} {2016})}\BibitemShut {NoStop}%
\bibitem [{\citenamefont {Levita}\ \emph {et~al.}(2015)\citenamefont {Levita},
  \citenamefont {Molinari}, \citenamefont {Polcar},\ and\ \citenamefont
  {Righi}}]{Levita-2015}%
  \BibitemOpen
  \bibfield  {author} {\bibinfo {author} {\bibfnamefont {G.}~\bibnamefont
  {Levita}}, \bibinfo {author} {\bibfnamefont {E.}~\bibnamefont {Molinari}},
  \bibinfo {author} {\bibfnamefont {T.}~\bibnamefont {Polcar}}, \ and\ \bibinfo
  {author} {\bibfnamefont {M.~C.}\ \bibnamefont {Righi}},\ }\href {\doibase
  10.1103/PhysRevB.92.085434} {\bibfield  {journal} {\bibinfo  {journal} {Phys.
  Rev. B}\ }\textbf {\bibinfo {volume} {92}},\ \bibinfo {pages} {085434}
  (\bibinfo {year} {2015})}\BibitemShut {NoStop}%
\bibitem [{\citenamefont {Kuwahara}\ \emph
  {et~al.}(2017{\natexlab{b}})\citenamefont {Kuwahara}, \citenamefont {Moras},\
  and\ \citenamefont {Moseler}}]{Stella-2017}%
  \BibitemOpen
  \bibfield  {author} {\bibinfo {author} {\bibfnamefont {T.}~\bibnamefont
  {Kuwahara}}, \bibinfo {author} {\bibfnamefont {G.}~\bibnamefont {Moras}}, \
  and\ \bibinfo {author} {\bibfnamefont {M.}~\bibnamefont {Moseler}},\ }\href
  {\doibase 10.1103/PhysRevLett.119.096101} {\bibfield  {journal} {\bibinfo
  {journal} {Phys. Rev. Lett.}\ }\textbf {\bibinfo {volume} {119}},\ \bibinfo
  {pages} {096101} (\bibinfo {year} {2017}{\natexlab{b}})}\BibitemShut
  {NoStop}%
\bibitem [{\citenamefont {Peeters}\ \emph {et~al.}(2020)\citenamefont
  {Peeters}, \citenamefont {Restuccia}, \citenamefont {Loehlé}, \citenamefont
  {Thiebaut},\ and\ \citenamefont {Righi}}]{Peeters-2020}%
  \BibitemOpen
  \bibfield  {author} {\bibinfo {author} {\bibfnamefont {S.}~\bibnamefont
  {Peeters}}, \bibinfo {author} {\bibfnamefont {P.}~\bibnamefont {Restuccia}},
  \bibinfo {author} {\bibfnamefont {S.}~\bibnamefont {Loehlé}}, \bibinfo
  {author} {\bibfnamefont {B.}~\bibnamefont {Thiebaut}}, \ and\ \bibinfo
  {author} {\bibfnamefont {M.~C.}\ \bibnamefont {Righi}},\ }\href {\doibase
  10.1021/acs.jpcc.0c02211} {\bibfield  {journal} {\bibinfo  {journal} {The
  Journal of Physical Chemistry C}\ }\textbf {\bibinfo {volume} {124}},\
  \bibinfo {pages} {13688} (\bibinfo {year} {2020})}\BibitemShut {NoStop}%
\bibitem [{\citenamefont {Le}\ \emph {et~al.}(2018)\citenamefont {Le},
  \citenamefont {Tieu}, \citenamefont {Zhu}, \citenamefont {Ta}, \citenamefont
  {Yu}, \citenamefont {Ta}, \citenamefont {Tran},\ and\ \citenamefont
  {Wan}}]{NVT-2018}%
  \BibitemOpen
  \bibfield  {author} {\bibinfo {author} {\bibfnamefont {M.~H.}\ \bibnamefont
  {Le}}, \bibinfo {author} {\bibfnamefont {A.~K.}\ \bibnamefont {Tieu}},
  \bibinfo {author} {\bibfnamefont {H.}~\bibnamefont {Zhu}}, \bibinfo {author}
  {\bibfnamefont {D.~T.}\ \bibnamefont {Ta}}, \bibinfo {author} {\bibfnamefont
  {H.}~\bibnamefont {Yu}}, \bibinfo {author} {\bibfnamefont {T.~T.~H.}\
  \bibnamefont {Ta}}, \bibinfo {author} {\bibfnamefont {V.~N.}\ \bibnamefont
  {Tran}}, \ and\ \bibinfo {author} {\bibfnamefont {S.}~\bibnamefont {Wan}},\
  }\href {\doibase 10.1039/C7CP08364E} {\bibfield  {journal} {\bibinfo
  {journal} {Phys. Chem. Chem. Phys.}\ }\textbf {\bibinfo {volume} {20}},\
  \bibinfo {pages} {7819} (\bibinfo {year} {2018})}\BibitemShut {NoStop}%
\bibitem [{\citenamefont {Ootani}\ \emph {et~al.}(2018)\citenamefont {Ootani},
  \citenamefont {Xu}, \citenamefont {Hatano},\ and\ \citenamefont
  {Kubo}}]{Kubo-2018}%
  \BibitemOpen
  \bibfield  {author} {\bibinfo {author} {\bibfnamefont {Y.}~\bibnamefont
  {Ootani}}, \bibinfo {author} {\bibfnamefont {J.}~\bibnamefont {Xu}}, \bibinfo
  {author} {\bibfnamefont {T.}~\bibnamefont {Hatano}}, \ and\ \bibinfo {author}
  {\bibfnamefont {M.}~\bibnamefont {Kubo}},\ }\href {\doibase
  10.1021/acs.jpcc.8b01953} {\bibfield  {journal} {\bibinfo  {journal} {The
  Journal of Physical Chemistry C}\ }\textbf {\bibinfo {volume} {122}},\
  \bibinfo {pages} {10459} (\bibinfo {year} {2018})}\BibitemShut {NoStop}%
\bibitem [{\citenamefont {Ramirez}\ \emph {et~al.}(2020)\citenamefont
  {Ramirez}, \citenamefont {Eryilmaz}, \citenamefont {Fatti}, \citenamefont
  {Righi}, \citenamefont {Wen},\ and\ \citenamefont
  {Erdemir}}]{Giulio-Erdemir}%
  \BibitemOpen
  \bibfield  {author} {\bibinfo {author} {\bibfnamefont {G.}~\bibnamefont
  {Ramirez}}, \bibinfo {author} {\bibfnamefont {O.~L.}\ \bibnamefont
  {Eryilmaz}}, \bibinfo {author} {\bibfnamefont {G.}~\bibnamefont {Fatti}},
  \bibinfo {author} {\bibfnamefont {M.~C.}\ \bibnamefont {Righi}}, \bibinfo
  {author} {\bibfnamefont {J.}~\bibnamefont {Wen}}, \ and\ \bibinfo {author}
  {\bibfnamefont {A.}~\bibnamefont {Erdemir}},\ }\href
  {https://doi.org/10.1021/acsanm.0c01527} {\bibfield  {journal} {\bibinfo
  {journal} {ACS Applied Nano Materials}\ }\textbf {\bibinfo {volume} {3}},\
  \bibinfo {pages} {8060} (\bibinfo {year} {2020})}\BibitemShut {NoStop}%
\bibitem [{\citenamefont {Melis}\ \emph {et~al.}(2014)\citenamefont {Melis},
  \citenamefont {Dettori}, \citenamefont {Vandermeulen},\ and\ \citenamefont
  {Colombo}}]{melis2014calculating}%
  \BibitemOpen
  \bibfield  {author} {\bibinfo {author} {\bibfnamefont {C.}~\bibnamefont
  {Melis}}, \bibinfo {author} {\bibfnamefont {R.}~\bibnamefont {Dettori}},
  \bibinfo {author} {\bibfnamefont {S.}~\bibnamefont {Vandermeulen}}, \ and\
  \bibinfo {author} {\bibfnamefont {L.}~\bibnamefont {Colombo}},\ }\href@noop
  {} {\bibfield  {journal} {\bibinfo  {journal} {The European Physical Journal
  B}\ }\textbf {\bibinfo {volume} {87}},\ \bibinfo {pages} {96} (\bibinfo
  {year} {2014})}\BibitemShut {NoStop}%
\bibitem [{\citenamefont {Kajita}\ \emph {et~al.}(2009)\citenamefont {Kajita},
  \citenamefont {Washizu},\ and\ \citenamefont {Ohmori}}]{kajita2009deep}%
  \BibitemOpen
  \bibfield  {author} {\bibinfo {author} {\bibfnamefont {S.}~\bibnamefont
  {Kajita}}, \bibinfo {author} {\bibfnamefont {H.}~\bibnamefont {Washizu}}, \
  and\ \bibinfo {author} {\bibfnamefont {T.}~\bibnamefont {Ohmori}},\
  }\href@noop {} {\bibfield  {journal} {\bibinfo  {journal} {EPL (Europhysics
  Letters)}\ }\textbf {\bibinfo {volume} {87}},\ \bibinfo {pages} {66002}
  (\bibinfo {year} {2009})}\BibitemShut {NoStop}%
\bibitem [{\citenamefont {Kajita}\ \emph {et~al.}(2010)\citenamefont {Kajita},
  \citenamefont {Washizu},\ and\ \citenamefont {Ohmori}}]{kajita2010approach}%
  \BibitemOpen
  \bibfield  {author} {\bibinfo {author} {\bibfnamefont {S.}~\bibnamefont
  {Kajita}}, \bibinfo {author} {\bibfnamefont {H.}~\bibnamefont {Washizu}}, \
  and\ \bibinfo {author} {\bibfnamefont {T.}~\bibnamefont {Ohmori}},\
  }\href@noop {} {\bibfield  {journal} {\bibinfo  {journal} {Physical Review
  B}\ }\textbf {\bibinfo {volume} {82}},\ \bibinfo {pages} {115424} (\bibinfo
  {year} {2010})}\BibitemShut {NoStop}%
\bibitem [{\citenamefont {Kajita}\ \emph {et~al.}(2012)\citenamefont {Kajita},
  \citenamefont {Washizu},\ and\ \citenamefont
  {Ohmori}}]{kajita2012simulation}%
  \BibitemOpen
  \bibfield  {author} {\bibinfo {author} {\bibfnamefont {S.}~\bibnamefont
  {Kajita}}, \bibinfo {author} {\bibfnamefont {H.}~\bibnamefont {Washizu}}, \
  and\ \bibinfo {author} {\bibfnamefont {T.}~\bibnamefont {Ohmori}},\
  }\href@noop {} {\bibfield  {journal} {\bibinfo  {journal} {Physical Review
  B}\ }\textbf {\bibinfo {volume} {86}},\ \bibinfo {pages} {075453} (\bibinfo
  {year} {2012})}\BibitemShut {NoStop}%
\bibitem [{\citenamefont {Zwanzig}(1960)}]{zwanzig1960collision}%
  \BibitemOpen
  \bibfield  {author} {\bibinfo {author} {\bibfnamefont {R.~W.}\ \bibnamefont
  {Zwanzig}},\ }\href@noop {} {\bibfield  {journal} {\bibinfo  {journal} {The
  Journal of Chemical Physics}\ }\textbf {\bibinfo {volume} {32}},\ \bibinfo
  {pages} {1173} (\bibinfo {year} {1960})}\BibitemShut {NoStop}%
\bibitem [{\citenamefont {Sokoloff}(1990)}]{sokoloff1990theory}%
  \BibitemOpen
  \bibfield  {author} {\bibinfo {author} {\bibfnamefont {J.}~\bibnamefont
  {Sokoloff}},\ }\href@noop {} {\bibfield  {journal} {\bibinfo  {journal}
  {Physical Review B}\ }\textbf {\bibinfo {volume} {42}},\ \bibinfo {pages}
  {760} (\bibinfo {year} {1990})}\BibitemShut {NoStop}%
\bibitem [{\citenamefont {Braun}\ \emph {et~al.}(2005)\citenamefont {Braun},
  \citenamefont {Peyrard}, \citenamefont {Bortolani}, \citenamefont
  {Franchini},\ and\ \citenamefont {Vanossi}}]{braun2005transition}%
  \BibitemOpen
  \bibfield  {author} {\bibinfo {author} {\bibfnamefont {O.}~\bibnamefont
  {Braun}}, \bibinfo {author} {\bibfnamefont {M.}~\bibnamefont {Peyrard}},
  \bibinfo {author} {\bibfnamefont {V.}~\bibnamefont {Bortolani}}, \bibinfo
  {author} {\bibfnamefont {A.}~\bibnamefont {Franchini}}, \ and\ \bibinfo
  {author} {\bibfnamefont {A.}~\bibnamefont {Vanossi}},\ }\href@noop {}
  {\bibfield  {journal} {\bibinfo  {journal} {Physical Review E}\ }\textbf
  {\bibinfo {volume} {72}},\ \bibinfo {pages} {056116} (\bibinfo {year}
  {2005})}\BibitemShut {NoStop}%
\bibitem [{\citenamefont {Campan{\'a}}\ and\ \citenamefont
  {M{\"u}ser}(2006)}]{campana2006practical}%
  \BibitemOpen
  \bibfield  {author} {\bibinfo {author} {\bibfnamefont {C.}~\bibnamefont
  {Campan{\'a}}}\ and\ \bibinfo {author} {\bibfnamefont {M.~H.}\ \bibnamefont
  {M{\"u}ser}},\ }\href@noop {} {\bibfield  {journal} {\bibinfo  {journal}
  {Physical Review B}\ }\textbf {\bibinfo {volume} {74}},\ \bibinfo {pages}
  {075420} (\bibinfo {year} {2006})}\BibitemShut {NoStop}%
\bibitem [{\citenamefont {Cai}\ \emph {et~al.}(2001)\citenamefont {Cai},
  \citenamefont {Bulatov}, \citenamefont {Chang}, \citenamefont {Li},\ and\
  \citenamefont {Yip}}]{cai2001anisotropic}%
  \BibitemOpen
  \bibfield  {author} {\bibinfo {author} {\bibfnamefont {W.}~\bibnamefont
  {Cai}}, \bibinfo {author} {\bibfnamefont {V.~V.}\ \bibnamefont {Bulatov}},
  \bibinfo {author} {\bibfnamefont {J.}~\bibnamefont {Chang}}, \bibinfo
  {author} {\bibfnamefont {J.}~\bibnamefont {Li}}, \ and\ \bibinfo {author}
  {\bibfnamefont {S.}~\bibnamefont {Yip}},\ }\href@noop {} {\bibfield
  {journal} {\bibinfo  {journal} {Physical Review Letters}\ }\textbf {\bibinfo
  {volume} {86}},\ \bibinfo {pages} {5727} (\bibinfo {year}
  {2001})}\BibitemShut {NoStop}%
\bibitem [{\citenamefont {Kajita}(2016)}]{kajita2016green}%
  \BibitemOpen
  \bibfield  {author} {\bibinfo {author} {\bibfnamefont {S.}~\bibnamefont
  {Kajita}},\ }\href@noop {} {\bibfield  {journal} {\bibinfo  {journal}
  {Physical Review E}\ }\textbf {\bibinfo {volume} {94}},\ \bibinfo {pages}
  {033301} (\bibinfo {year} {2016})}\BibitemShut {NoStop}%
\bibitem [{\citenamefont {Monti}\ \emph {et~al.}(2021)\citenamefont {Monti},
  \citenamefont {Pastewka},\ and\ \citenamefont {Robbins}}]{Monti-2021}%
  \BibitemOpen
  \bibfield  {author} {\bibinfo {author} {\bibfnamefont {J.~M.}\ \bibnamefont
  {Monti}}, \bibinfo {author} {\bibfnamefont {L.}~\bibnamefont {Pastewka}}, \
  and\ \bibinfo {author} {\bibfnamefont {M.~O.}\ \bibnamefont {Robbins}},\
  }\href {\doibase 10.1103/PhysRevE.103.053305} {\bibfield  {journal} {\bibinfo
   {journal} {Phys. Rev. E}\ }\textbf {\bibinfo {volume} {103}},\ \bibinfo
  {pages} {053305} (\bibinfo {year} {2021})}\BibitemShut {NoStop}%
\bibitem [{\citenamefont {Talbot}(1979)}]{talbot1979accurate}%
  \BibitemOpen
  \bibfield  {author} {\bibinfo {author} {\bibfnamefont {A.}~\bibnamefont
  {Talbot}},\ }\href@noop {} {\bibfield  {journal} {\bibinfo  {journal} {IMA
  Journal of Applied Mathematics}\ }\textbf {\bibinfo {volume} {23}},\ \bibinfo
  {pages} {97} (\bibinfo {year} {1979})}\BibitemShut {NoStop}%
\bibitem [{\citenamefont {Lubich}\ and\ \citenamefont
  {Sch{\"a}dle}(2002)}]{lubich2002fast}%
  \BibitemOpen
  \bibfield  {author} {\bibinfo {author} {\bibfnamefont {C.}~\bibnamefont
  {Lubich}}\ and\ \bibinfo {author} {\bibfnamefont {A.}~\bibnamefont
  {Sch{\"a}dle}},\ }\href@noop {} {\bibfield  {journal} {\bibinfo  {journal}
  {SIAM Journal on Scientific Computing}\ }\textbf {\bibinfo {volume} {24}},\
  \bibinfo {pages} {161} (\bibinfo {year} {2002})}\BibitemShut {NoStop}%
\bibitem [{\citenamefont {Capobianco}\ \emph {et~al.}(2007)\citenamefont
  {Capobianco}, \citenamefont {Conte}, \citenamefont {Prete},\ and\
  \citenamefont {Russo}}]{capobianco2007fast}%
  \BibitemOpen
  \bibfield  {author} {\bibinfo {author} {\bibfnamefont {G.}~\bibnamefont
  {Capobianco}}, \bibinfo {author} {\bibfnamefont {D.}~\bibnamefont {Conte}},
  \bibinfo {author} {\bibfnamefont {I.~D.}\ \bibnamefont {Prete}}, \ and\
  \bibinfo {author} {\bibfnamefont {E.}~\bibnamefont {Russo}},\ }\href@noop {}
  {\bibfield  {journal} {\bibinfo  {journal} {BIT Numerical Mathematics}\
  }\textbf {\bibinfo {volume} {47}},\ \bibinfo {pages} {259} (\bibinfo {year}
  {2007})}\BibitemShut {NoStop}%
\bibitem [{\citenamefont {Prete}(2006)}]{phdPrete}%
  \BibitemOpen
  \bibfield  {author} {\bibinfo {author} {\bibfnamefont {I.~D.}\ \bibnamefont
  {Prete}},\ }\emph {\bibinfo {title} {Efficient numerical methods for Volterra
  integral equations of Hammerstein type, Doctoral thesis}},\ \href@noop {}
  {Ph.D. thesis},\ \bibinfo  {school} {Universita degli Studi di Napoli
  Federico II} (\bibinfo {year} {2006})\BibitemShut {NoStop}%
\bibitem [{\citenamefont {Swart}(2003)}]{swart2003addremove}%
  \BibitemOpen
  \bibfield  {author} {\bibinfo {author} {\bibfnamefont {M.}~\bibnamefont
  {Swart}},\ }\href@noop {} {\bibfield  {journal} {\bibinfo  {journal}
  {International Journal of Quantum Chemistry}\ }\textbf {\bibinfo {volume}
  {91}},\ \bibinfo {pages} {177} (\bibinfo {year} {2003})}\BibitemShut
  {NoStop}%
\bibitem [{\citenamefont {Farlow}(1993)}]{StanleyJFarlow}%
  \BibitemOpen
  \bibfield  {author} {\bibinfo {author} {\bibfnamefont {S.~J.}\ \bibnamefont
  {Farlow}},\ }\href@noop {} {\emph {\bibinfo {title} {Partial Differential
  Equations for Scientists and Engineers}}}\ (\bibinfo  {publisher} {Dover
  Publication},\ \bibinfo {year} {1993})\BibitemShut {NoStop}%
\bibitem [{\citenamefont {Berkowitz}\ \emph {et~al.}(1983)\citenamefont
  {Berkowitz}, \citenamefont {Morgan},\ and\ \citenamefont
  {McCammon}}]{berkowitz1983generalized}%
  \BibitemOpen
  \bibfield  {author} {\bibinfo {author} {\bibfnamefont {M.}~\bibnamefont
  {Berkowitz}}, \bibinfo {author} {\bibfnamefont {J.}~\bibnamefont {Morgan}}, \
  and\ \bibinfo {author} {\bibfnamefont {J.~A.}\ \bibnamefont {McCammon}},\
  }\href@noop {} {\bibfield  {journal} {\bibinfo  {journal} {The Journal of
  Chemical Physics}\ }\textbf {\bibinfo {volume} {78}},\ \bibinfo {pages}
  {3256} (\bibinfo {year} {1983})}\BibitemShut {NoStop}%
\bibitem [{\citenamefont {Venugopalan}\ \emph {et~al.}(2017)\citenamefont
  {Venugopalan}, \citenamefont {M{\"u}ser},\ and\ \citenamefont
  {Nicola}}]{venugopalan2017green}%
  \BibitemOpen
  \bibfield  {author} {\bibinfo {author} {\bibfnamefont {S.~P.}\ \bibnamefont
  {Venugopalan}}, \bibinfo {author} {\bibfnamefont {M.~H.}\ \bibnamefont
  {M{\"u}ser}}, \ and\ \bibinfo {author} {\bibfnamefont {L.}~\bibnamefont
  {Nicola}},\ }\href@noop {} {\bibfield  {journal} {\bibinfo  {journal}
  {Modelling and Simulation in Materials Science and Engineering}\ }\textbf
  {\bibinfo {volume} {25}},\ \bibinfo {pages} {065018} (\bibinfo {year}
  {2017})}\BibitemShut {NoStop}%
\bibitem [{\citenamefont {Otani}\ and\ \citenamefont
  {Sugino}(2006)}]{otani2006first}%
  \BibitemOpen
  \bibfield  {author} {\bibinfo {author} {\bibfnamefont {M.}~\bibnamefont
  {Otani}}\ and\ \bibinfo {author} {\bibfnamefont {O.}~\bibnamefont {Sugino}},\
  }\href@noop {} {\bibfield  {journal} {\bibinfo  {journal} {Physical Review
  B}\ }\textbf {\bibinfo {volume} {73}},\ \bibinfo {pages} {115407} (\bibinfo
  {year} {2006})}\BibitemShut {NoStop}%
\bibitem [{\citenamefont {Ohba}\ and\ \citenamefont
  {Ogata}(2020)}]{ohba2020large}%
  \BibitemOpen
  \bibfield  {author} {\bibinfo {author} {\bibfnamefont {N.}~\bibnamefont
  {Ohba}}\ and\ \bibinfo {author} {\bibfnamefont {S.}~\bibnamefont {Ogata}},\
  }in\ \href@noop {} {\emph {\bibinfo {booktitle} {Multiscale Simulations for
  Electrochemical Devices}}}\ (\bibinfo  {publisher} {Jenny Stanford
  Publishing},\ \bibinfo {year} {2020})\ pp.\ \bibinfo {pages}
  {147--194}\BibitemShut {NoStop}%
\bibitem [{\citenamefont {Giannozzi}\ \emph {et~al.}(1991)\citenamefont
  {Giannozzi}, \citenamefont {De~Gironcoli}, \citenamefont {Pavone},\ and\
  \citenamefont {Baroni}}]{qephonon}%
  \BibitemOpen
  \bibfield  {author} {\bibinfo {author} {\bibfnamefont {P.}~\bibnamefont
  {Giannozzi}}, \bibinfo {author} {\bibfnamefont {S.}~\bibnamefont
  {De~Gironcoli}}, \bibinfo {author} {\bibfnamefont {P.}~\bibnamefont
  {Pavone}}, \ and\ \bibinfo {author} {\bibfnamefont {S.}~\bibnamefont
  {Baroni}},\ }\href@noop {} {\bibfield  {journal} {\bibinfo  {journal}
  {Physical Review B}\ }\textbf {\bibinfo {volume} {43}},\ \bibinfo {pages}
  {7231} (\bibinfo {year} {1991})}\BibitemShut {NoStop}%
\bibitem [{\citenamefont {Giannozzi}\ \emph {et~al.}(2009)\citenamefont
  {Giannozzi}, \citenamefont {Baroni}, \citenamefont {Bonini}, \citenamefont
  {Calandra}, \citenamefont {Car}, \citenamefont {Cavazzoni}, \citenamefont
  {Ceresoli}, \citenamefont {Chiarotti}, \citenamefont {Cococcioni},
  \citenamefont {Dabo} \emph {et~al.}}]{qe}%
  \BibitemOpen
  \bibfield  {author} {\bibinfo {author} {\bibfnamefont {P.}~\bibnamefont
  {Giannozzi}}, \bibinfo {author} {\bibfnamefont {S.}~\bibnamefont {Baroni}},
  \bibinfo {author} {\bibfnamefont {N.}~\bibnamefont {Bonini}}, \bibinfo
  {author} {\bibfnamefont {M.}~\bibnamefont {Calandra}}, \bibinfo {author}
  {\bibfnamefont {R.}~\bibnamefont {Car}}, \bibinfo {author} {\bibfnamefont
  {C.}~\bibnamefont {Cavazzoni}}, \bibinfo {author} {\bibfnamefont
  {D.}~\bibnamefont {Ceresoli}}, \bibinfo {author} {\bibfnamefont {G.~L.}\
  \bibnamefont {Chiarotti}}, \bibinfo {author} {\bibfnamefont {M.}~\bibnamefont
  {Cococcioni}}, \bibinfo {author} {\bibfnamefont {I.}~\bibnamefont {Dabo}},
  \emph {et~al.},\ }\href@noop {} {\bibfield  {journal} {\bibinfo  {journal}
  {Journal of physics: Condensed matter}\ }\textbf {\bibinfo {volume} {21}},\
  \bibinfo {pages} {395502} (\bibinfo {year} {2009})}\BibitemShut {NoStop}%
\bibitem [{\citenamefont {Giannozzi}\ \emph {et~al.}(2017)\citenamefont
  {Giannozzi}, \citenamefont {Andreussi}, \citenamefont {Brumme}, \citenamefont
  {Bunau}, \citenamefont {Nardelli}, \citenamefont {Calandra}, \citenamefont
  {Car}, \citenamefont {Cavazzoni}, \citenamefont {Ceresoli}, \citenamefont
  {Cococcioni}, \citenamefont {Colonna}, \citenamefont {Carnimeo},
  \citenamefont {Corso}, \citenamefont {de~Gironcoli}, \citenamefont {Delugas},
  \citenamefont {DiStasio}, \citenamefont {Ferretti}, \citenamefont {Floris},
  \citenamefont {Fratesi}, \citenamefont {Fugallo}, \citenamefont {Gebauer},
  \citenamefont {Gerstmann}, \citenamefont {Giustino}, \citenamefont {Gorni},
  \citenamefont {Jia}, \citenamefont {Kawamura}, \citenamefont {Ko},
  \citenamefont {Kokalj}, \citenamefont {Kü{\c{c}}ükbenli}, \citenamefont
  {Lazzeri}, \citenamefont {Marsili}, \citenamefont {Marzari}, \citenamefont
  {Mauri}, \citenamefont {Nguyen}, \citenamefont {Nguyen}, \citenamefont {de-la
  Roza}, \citenamefont {Paulatto}, \citenamefont {Ponc{\'{e}}}, \citenamefont
  {Rocca}, \citenamefont {Sabatini}, \citenamefont {Santra}, \citenamefont
  {Schlipf}, \citenamefont {Seitsonen}, \citenamefont {Smogunov}, \citenamefont
  {Timrov}, \citenamefont {Thonhauser}, \citenamefont {Umari}, \citenamefont
  {Vast}, \citenamefont {Wu},\ and\ \citenamefont {Baroni}}]{qe2}%
  \BibitemOpen
  \bibfield  {author} {\bibinfo {author} {\bibfnamefont {P.}~\bibnamefont
  {Giannozzi}}, \bibinfo {author} {\bibfnamefont {O.}~\bibnamefont
  {Andreussi}}, \bibinfo {author} {\bibfnamefont {T.}~\bibnamefont {Brumme}},
  \bibinfo {author} {\bibfnamefont {O.}~\bibnamefont {Bunau}}, \bibinfo
  {author} {\bibfnamefont {M.~B.}\ \bibnamefont {Nardelli}}, \bibinfo {author}
  {\bibfnamefont {M.}~\bibnamefont {Calandra}}, \bibinfo {author}
  {\bibfnamefont {R.}~\bibnamefont {Car}}, \bibinfo {author} {\bibfnamefont
  {C.}~\bibnamefont {Cavazzoni}}, \bibinfo {author} {\bibfnamefont
  {D.}~\bibnamefont {Ceresoli}}, \bibinfo {author} {\bibfnamefont
  {M.}~\bibnamefont {Cococcioni}}, \bibinfo {author} {\bibfnamefont
  {N.}~\bibnamefont {Colonna}}, \bibinfo {author} {\bibfnamefont
  {I.}~\bibnamefont {Carnimeo}}, \bibinfo {author} {\bibfnamefont {A.~D.}\
  \bibnamefont {Corso}}, \bibinfo {author} {\bibfnamefont {S.}~\bibnamefont
  {de~Gironcoli}}, \bibinfo {author} {\bibfnamefont {P.}~\bibnamefont
  {Delugas}}, \bibinfo {author} {\bibfnamefont {R.~A.}\ \bibnamefont
  {DiStasio}}, \bibinfo {author} {\bibfnamefont {A.}~\bibnamefont {Ferretti}},
  \bibinfo {author} {\bibfnamefont {A.}~\bibnamefont {Floris}}, \bibinfo
  {author} {\bibfnamefont {G.}~\bibnamefont {Fratesi}}, \bibinfo {author}
  {\bibfnamefont {G.}~\bibnamefont {Fugallo}}, \bibinfo {author} {\bibfnamefont
  {R.}~\bibnamefont {Gebauer}}, \bibinfo {author} {\bibfnamefont
  {U.}~\bibnamefont {Gerstmann}}, \bibinfo {author} {\bibfnamefont
  {F.}~\bibnamefont {Giustino}}, \bibinfo {author} {\bibfnamefont
  {T.}~\bibnamefont {Gorni}}, \bibinfo {author} {\bibfnamefont
  {J.}~\bibnamefont {Jia}}, \bibinfo {author} {\bibfnamefont {M.}~\bibnamefont
  {Kawamura}}, \bibinfo {author} {\bibfnamefont {H.-Y.}\ \bibnamefont {Ko}},
  \bibinfo {author} {\bibfnamefont {A.}~\bibnamefont {Kokalj}}, \bibinfo
  {author} {\bibfnamefont {E.}~\bibnamefont {Kü{\c{c}}ükbenli}}, \bibinfo
  {author} {\bibfnamefont {M.}~\bibnamefont {Lazzeri}}, \bibinfo {author}
  {\bibfnamefont {M.}~\bibnamefont {Marsili}}, \bibinfo {author} {\bibfnamefont
  {N.}~\bibnamefont {Marzari}}, \bibinfo {author} {\bibfnamefont
  {F.}~\bibnamefont {Mauri}}, \bibinfo {author} {\bibfnamefont {N.~L.}\
  \bibnamefont {Nguyen}}, \bibinfo {author} {\bibfnamefont {H.-V.}\
  \bibnamefont {Nguyen}}, \bibinfo {author} {\bibfnamefont {A.~O.}\
  \bibnamefont {de-la Roza}}, \bibinfo {author} {\bibfnamefont
  {L.}~\bibnamefont {Paulatto}}, \bibinfo {author} {\bibfnamefont
  {S.}~\bibnamefont {Ponc{\'{e}}}}, \bibinfo {author} {\bibfnamefont
  {D.}~\bibnamefont {Rocca}}, \bibinfo {author} {\bibfnamefont
  {R.}~\bibnamefont {Sabatini}}, \bibinfo {author} {\bibfnamefont
  {B.}~\bibnamefont {Santra}}, \bibinfo {author} {\bibfnamefont
  {M.}~\bibnamefont {Schlipf}}, \bibinfo {author} {\bibfnamefont {A.~P.}\
  \bibnamefont {Seitsonen}}, \bibinfo {author} {\bibfnamefont {A.}~\bibnamefont
  {Smogunov}}, \bibinfo {author} {\bibfnamefont {I.}~\bibnamefont {Timrov}},
  \bibinfo {author} {\bibfnamefont {T.}~\bibnamefont {Thonhauser}}, \bibinfo
  {author} {\bibfnamefont {P.}~\bibnamefont {Umari}}, \bibinfo {author}
  {\bibfnamefont {N.}~\bibnamefont {Vast}}, \bibinfo {author} {\bibfnamefont
  {X.}~\bibnamefont {Wu}}, \ and\ \bibinfo {author} {\bibfnamefont
  {S.}~\bibnamefont {Baroni}},\ }\href {\doibase 10.1088/1361-648x/aa8f79}
  {\bibfield  {journal} {\bibinfo  {journal} {Journal of Physics: Condensed
  Matter}\ }\textbf {\bibinfo {volume} {29}},\ \bibinfo {pages} {465901}
  (\bibinfo {year} {2017})}\BibitemShut {NoStop}%
\bibitem [{\citenamefont {Giannozzi}\ \emph {et~al.}(2020)\citenamefont
  {Giannozzi}, \citenamefont {Baseggio}, \citenamefont {Bonfà}, \citenamefont
  {Brunato}, \citenamefont {Car}, \citenamefont {Carnimeo}, \citenamefont
  {Cavazzoni}, \citenamefont {de~Gironcoli}, \citenamefont {Delugas},
  \citenamefont {Ferrari~Ruffino}, \citenamefont {Ferretti}, \citenamefont
  {Marzari}, \citenamefont {Timrov}, \citenamefont {Urru},\ and\ \citenamefont
  {Baroni}}]{qe3}%
  \BibitemOpen
  \bibfield  {author} {\bibinfo {author} {\bibfnamefont {P.}~\bibnamefont
  {Giannozzi}}, \bibinfo {author} {\bibfnamefont {O.}~\bibnamefont {Baseggio}},
  \bibinfo {author} {\bibfnamefont {P.}~\bibnamefont {Bonfà}}, \bibinfo
  {author} {\bibfnamefont {D.}~\bibnamefont {Brunato}}, \bibinfo {author}
  {\bibfnamefont {R.}~\bibnamefont {Car}}, \bibinfo {author} {\bibfnamefont
  {I.}~\bibnamefont {Carnimeo}}, \bibinfo {author} {\bibfnamefont
  {C.}~\bibnamefont {Cavazzoni}}, \bibinfo {author} {\bibfnamefont
  {S.}~\bibnamefont {de~Gironcoli}}, \bibinfo {author} {\bibfnamefont
  {P.}~\bibnamefont {Delugas}}, \bibinfo {author} {\bibfnamefont
  {F.}~\bibnamefont {Ferrari~Ruffino}}, \bibinfo {author} {\bibfnamefont
  {A.}~\bibnamefont {Ferretti}}, \bibinfo {author} {\bibfnamefont
  {N.}~\bibnamefont {Marzari}}, \bibinfo {author} {\bibfnamefont
  {I.}~\bibnamefont {Timrov}}, \bibinfo {author} {\bibfnamefont
  {A.}~\bibnamefont {Urru}}, \ and\ \bibinfo {author} {\bibfnamefont
  {S.}~\bibnamefont {Baroni}},\ }\href {\doibase 10.1063/5.0005082} {\bibfield
  {journal} {\bibinfo  {journal} {The Journal of Chemical Physics}\ }\textbf
  {\bibinfo {volume} {152}},\ \bibinfo {pages} {154105} (\bibinfo {year}
  {2020})}\BibitemShut {NoStop}%
\bibitem [{\citenamefont {Perdew}\ \emph {et~al.}(1996)\citenamefont {Perdew},
  \citenamefont {Burke},\ and\ \citenamefont {Ernzerhof}}]{pbe}%
  \BibitemOpen
  \bibfield  {author} {\bibinfo {author} {\bibfnamefont {J.~P.}\ \bibnamefont
  {Perdew}}, \bibinfo {author} {\bibfnamefont {K.}~\bibnamefont {Burke}}, \
  and\ \bibinfo {author} {\bibfnamefont {M.}~\bibnamefont {Ernzerhof}},\
  }\href@noop {} {\bibfield  {journal} {\bibinfo  {journal} {Physical review
  letters}\ }\textbf {\bibinfo {volume} {77}},\ \bibinfo {pages} {3865}
  (\bibinfo {year} {1996})}\BibitemShut {NoStop}%
\bibitem [{\citenamefont {Vanderbilt}(1990)}]{vanderbilt}%
  \BibitemOpen
  \bibfield  {author} {\bibinfo {author} {\bibfnamefont {D.}~\bibnamefont
  {Vanderbilt}},\ }\href@noop {} {\bibfield  {journal} {\bibinfo  {journal}
  {Physical review B}\ }\textbf {\bibinfo {volume} {41}},\ \bibinfo {pages}
  {7892} (\bibinfo {year} {1990})}\BibitemShut {NoStop}%
\bibitem [{\citenamefont {Kubo}(1966)}]{kubo-1966}%
  \BibitemOpen
  \bibfield  {author} {\bibinfo {author} {\bibfnamefont {R.}~\bibnamefont
  {Kubo}},\ }\href {\doibase 10.1088/0034-4885/29/1/306} {\ \textbf {\bibinfo
  {volume} {29}},\ \bibinfo {pages} {255} (\bibinfo {year} {1966})}\BibitemShut
  {NoStop}%
\bibitem [{\citenamefont {Tabor}(1979)}]{Tabor-1979}%
  \BibitemOpen
  \bibfield  {author} {\bibinfo {author} {\bibfnamefont {D.}~\bibnamefont
  {Tabor}},\ }\href {https://ci.nii.ac.jp/naid/10023980836/en/} {\bibfield
  {journal} {\bibinfo  {journal} {The Properties of Diamond}\ }\textbf
  {\bibinfo {volume} {325}} (\bibinfo {year} {1979})}\BibitemShut {NoStop}%
\bibitem [{\citenamefont {Hayward}(1991)}]{Hayward-1991}%
  \BibitemOpen
  \bibfield  {author} {\bibinfo {author} {\bibfnamefont {I.}~\bibnamefont
  {Hayward}},\ }\href {\doibase https://doi.org/10.1016/0257-8972(91)90116-E}
  {\bibfield  {journal} {\bibinfo  {journal} {Surface and Coatings Technology}\
  }\textbf {\bibinfo {volume} {49}},\ \bibinfo {pages} {554 } (\bibinfo {year}
  {1991})}\BibitemShut {NoStop}%
\bibitem [{\citenamefont {Erdemir}\ and\ \citenamefont
  {Martin}(2018)}]{Erdermir-2018}%
  \BibitemOpen
  \bibfield  {author} {\bibinfo {author} {\bibfnamefont {A.}~\bibnamefont
  {Erdemir}}\ and\ \bibinfo {author} {\bibfnamefont {J.~M.}\ \bibnamefont
  {Martin}},\ }\href {\doibase 10.1016/j.cossms.2018.11.003} {\bibfield
  {journal} {\bibinfo  {journal} {Current Opinion in Solid State and Materials
  Science}\ }\textbf {\bibinfo {volume} {22}},\ \bibinfo {pages} {243 }
  (\bibinfo {year} {2018})}\BibitemShut {NoStop}%
\bibitem [{\citenamefont {Samuels}\ and\ \citenamefont
  {Wilks}(1988)}]{Samuels-1988}%
  \BibitemOpen
  \bibfield  {author} {\bibinfo {author} {\bibfnamefont {B.}~\bibnamefont
  {Samuels}}\ and\ \bibinfo {author} {\bibfnamefont {J.}~\bibnamefont
  {Wilks}},\ }\href {\doibase 10.1007/BF00547459} {\bibfield  {journal}
  {\bibinfo  {journal} {Journal of Materials Science}\ }\textbf {\bibinfo
  {volume} {23}},\ \bibinfo {pages} {2846} (\bibinfo {year}
  {1988})}\BibitemShut {NoStop}%
\bibitem [{\citenamefont {Feng}\ and\ \citenamefont {Field}(1992)}]{Feng-1992}%
  \BibitemOpen
  \bibfield  {author} {\bibinfo {author} {\bibfnamefont {Z.}~\bibnamefont
  {Feng}}\ and\ \bibinfo {author} {\bibfnamefont {J.~E.}\ \bibnamefont
  {Field}},\ }\href {\doibase 10.1088/0022-3727/25/1a/007} {\bibfield
  {journal} {\bibinfo  {journal} {Journal of Physics D: Applied Physics}\
  }\textbf {\bibinfo {volume} {25}},\ \bibinfo {pages} {A33} (\bibinfo {year}
  {1992})}\BibitemShut {NoStop}%
\bibitem [{\citenamefont {Germann}\ \emph {et~al.}(1993)\citenamefont
  {Germann}, \citenamefont {Cohen}, \citenamefont {Neubauer}, \citenamefont
  {McClelland}, \citenamefont {Seki},\ and\ \citenamefont
  {Coulman}}]{Germann-1993}%
  \BibitemOpen
  \bibfield  {author} {\bibinfo {author} {\bibfnamefont {G.~J.}\ \bibnamefont
  {Germann}}, \bibinfo {author} {\bibfnamefont {S.~R.}\ \bibnamefont {Cohen}},
  \bibinfo {author} {\bibfnamefont {G.}~\bibnamefont {Neubauer}}, \bibinfo
  {author} {\bibfnamefont {G.~M.}\ \bibnamefont {McClelland}}, \bibinfo
  {author} {\bibfnamefont {H.}~\bibnamefont {Seki}}, \ and\ \bibinfo {author}
  {\bibfnamefont {D.}~\bibnamefont {Coulman}},\ }\href {\doibase
  10.1063/1.353878} {\bibfield  {journal} {\bibinfo  {journal} {Journal of
  Applied Physics}\ }\textbf {\bibinfo {volume} {73}},\ \bibinfo {pages} {163}
  (\bibinfo {year} {1993})}\BibitemShut {NoStop}%
\bibitem [{\citenamefont {Ajikumar}\ \emph {et~al.}(2019)\citenamefont
  {Ajikumar}, \citenamefont {Ganesan}, \citenamefont {Kumar}, \citenamefont
  {Ravindran}, \citenamefont {Kalavathi},\ and\ \citenamefont
  {Kamruddin}}]{Ajikumar-2019}%
  \BibitemOpen
  \bibfield  {author} {\bibinfo {author} {\bibfnamefont {P.}~\bibnamefont
  {Ajikumar}}, \bibinfo {author} {\bibfnamefont {K.}~\bibnamefont {Ganesan}},
  \bibinfo {author} {\bibfnamefont {N.}~\bibnamefont {Kumar}}, \bibinfo
  {author} {\bibfnamefont {T.}~\bibnamefont {Ravindran}}, \bibinfo {author}
  {\bibfnamefont {S.}~\bibnamefont {Kalavathi}}, \ and\ \bibinfo {author}
  {\bibfnamefont {M.}~\bibnamefont {Kamruddin}},\ }\href {\doibase
  10.1016/j.apsusc.2018.10.265} {\bibfield  {journal} {\bibinfo  {journal}
  {Applied Surface Science}\ }\textbf {\bibinfo {volume} {469}},\ \bibinfo
  {pages} {10 } (\bibinfo {year} {2019})}\BibitemShut {NoStop}%
\bibitem [{\citenamefont {Konicek}\ \emph {et~al.}(2008)\citenamefont
  {Konicek}, \citenamefont {Grierson}, \citenamefont {Gilbert}, \citenamefont
  {Sawyer}, \citenamefont {Sumant},\ and\ \citenamefont
  {Carpick}}]{Konicek-2008}%
  \BibitemOpen
  \bibfield  {author} {\bibinfo {author} {\bibfnamefont {A.~R.}\ \bibnamefont
  {Konicek}}, \bibinfo {author} {\bibfnamefont {D.~S.}\ \bibnamefont
  {Grierson}}, \bibinfo {author} {\bibfnamefont {P.~U. P.~A.}\ \bibnamefont
  {Gilbert}}, \bibinfo {author} {\bibfnamefont {W.~G.}\ \bibnamefont {Sawyer}},
  \bibinfo {author} {\bibfnamefont {A.~V.}\ \bibnamefont {Sumant}}, \ and\
  \bibinfo {author} {\bibfnamefont {R.~W.}\ \bibnamefont {Carpick}},\ }\href
  {\doibase 10.1103/PhysRevLett.100.235502} {\bibfield  {journal} {\bibinfo
  {journal} {Phys. Rev. Lett.}\ }\textbf {\bibinfo {volume} {100}},\ \bibinfo
  {pages} {235502} (\bibinfo {year} {2008})}\BibitemShut {NoStop}%
\bibitem [{\citenamefont {Wang}\ \emph {et~al.}(2013)\citenamefont {Wang},
  \citenamefont {Wang}, \citenamefont {Li}, \citenamefont {Sun}, \citenamefont
  {Yuan},\ and\ \citenamefont {Jia}}]{Wang-2013}%
  \BibitemOpen
  \bibfield  {author} {\bibinfo {author} {\bibfnamefont {J.}~\bibnamefont
  {Wang}}, \bibinfo {author} {\bibfnamefont {F.}~\bibnamefont {Wang}}, \bibinfo
  {author} {\bibfnamefont {J.}~\bibnamefont {Li}}, \bibinfo {author}
  {\bibfnamefont {Q.}~\bibnamefont {Sun}}, \bibinfo {author} {\bibfnamefont
  {P.}~\bibnamefont {Yuan}}, \ and\ \bibinfo {author} {\bibfnamefont
  {Y.}~\bibnamefont {Jia}},\ }\href {\doibase 10.1016/j.susc.2012.09.016}
  {\bibfield  {journal} {\bibinfo  {journal} {Surface Science}\ }\textbf
  {\bibinfo {volume} {608}},\ \bibinfo {pages} {74 } (\bibinfo {year}
  {2013})}\BibitemShut {NoStop}%
\bibitem [{\citenamefont {Cui}\ \emph {et~al.}(2014)\citenamefont {Cui},
  \citenamefont {Lu},\ and\ \citenamefont {Wang}}]{Cui-2014}%
  \BibitemOpen
  \bibfield  {author} {\bibinfo {author} {\bibfnamefont {L.}~\bibnamefont
  {Cui}}, \bibinfo {author} {\bibfnamefont {Z.}~\bibnamefont {Lu}}, \ and\
  \bibinfo {author} {\bibfnamefont {L.}~\bibnamefont {Wang}},\ }\href {\doibase
  10.1016/j.carbon.2013.08.065} {\bibfield  {journal} {\bibinfo  {journal}
  {Carbon}\ }\textbf {\bibinfo {volume} {66}},\ \bibinfo {pages} {259 }
  (\bibinfo {year} {2014})}\BibitemShut {NoStop}%
\bibitem [{\citenamefont {Zilibotti}\ \emph {et~al.}(2009)\citenamefont
  {Zilibotti}, \citenamefont {Righi},\ and\ \citenamefont
  {Ferrario}}]{Zilibotti-2009}%
  \BibitemOpen
  \bibfield  {author} {\bibinfo {author} {\bibfnamefont {G.}~\bibnamefont
  {Zilibotti}}, \bibinfo {author} {\bibfnamefont {M.~C.}\ \bibnamefont
  {Righi}}, \ and\ \bibinfo {author} {\bibfnamefont {M.}~\bibnamefont
  {Ferrario}},\ }\href {\doibase 10.1103/PhysRevB.79.075420} {\bibfield
  {journal} {\bibinfo  {journal} {Phys. Rev. B}\ }\textbf {\bibinfo {volume}
  {79}},\ \bibinfo {pages} {075420} (\bibinfo {year} {2009})}\BibitemShut
  {NoStop}%
\bibitem [{\citenamefont {Zilibotti}\ and\ \citenamefont
  {Righi}(2011)}]{Zilibotti-2011}%
  \BibitemOpen
  \bibfield  {author} {\bibinfo {author} {\bibfnamefont {G.}~\bibnamefont
  {Zilibotti}}\ and\ \bibinfo {author} {\bibfnamefont {M.~C.}\ \bibnamefont
  {Righi}},\ }\href {\doibase 10.1021/la200783a} {\bibfield  {journal}
  {\bibinfo  {journal} {Langmuir}\ }\textbf {\bibinfo {volume} {27}},\ \bibinfo
  {pages} {6862} (\bibinfo {year} {2011})},\ \bibinfo {note} {pMID:
  21545120}\BibitemShut {NoStop}%
\bibitem [{\citenamefont {De~Barros~Bouchet}\ \emph {et~al.}(2012)\citenamefont
  {De~Barros~Bouchet}, \citenamefont {Zilibotti}, \citenamefont {Matta},
  \citenamefont {Righi}, \citenamefont {Vandenbulcke}, \citenamefont {Vacher},\
  and\ \citenamefont {Martin}}]{Bouchet-2012}%
  \BibitemOpen
  \bibfield  {author} {\bibinfo {author} {\bibfnamefont {M.-I.}\ \bibnamefont
  {De~Barros~Bouchet}}, \bibinfo {author} {\bibfnamefont {G.}~\bibnamefont
  {Zilibotti}}, \bibinfo {author} {\bibfnamefont {C.}~\bibnamefont {Matta}},
  \bibinfo {author} {\bibfnamefont {M.~C.}\ \bibnamefont {Righi}}, \bibinfo
  {author} {\bibfnamefont {L.}~\bibnamefont {Vandenbulcke}}, \bibinfo {author}
  {\bibfnamefont {B.}~\bibnamefont {Vacher}}, \ and\ \bibinfo {author}
  {\bibfnamefont {J.-M.}\ \bibnamefont {Martin}},\ }\href {\doibase
  10.1021/jp211322s} {\bibfield  {journal} {\bibinfo  {journal} {The Journal of
  Physical Chemistry C}\ }\textbf {\bibinfo {volume} {116}},\ \bibinfo {pages}
  {6966} (\bibinfo {year} {2012})}\BibitemShut {NoStop}%
\bibitem [{\citenamefont {Kuwahara}\ \emph
  {et~al.}(2017{\natexlab{c}})\citenamefont {Kuwahara}, \citenamefont {Moras},\
  and\ \citenamefont {Moseler}}]{Kuwahara-2017}%
  \BibitemOpen
  \bibfield  {author} {\bibinfo {author} {\bibfnamefont {T.}~\bibnamefont
  {Kuwahara}}, \bibinfo {author} {\bibfnamefont {G.}~\bibnamefont {Moras}}, \
  and\ \bibinfo {author} {\bibfnamefont {M.}~\bibnamefont {Moseler}},\ }\href
  {\doibase 10.1103/PhysRevLett.119.096101} {\bibfield  {journal} {\bibinfo
  {journal} {Phys. Rev. Lett.}\ }\textbf {\bibinfo {volume} {119}},\ \bibinfo
  {pages} {096101} (\bibinfo {year} {2017}{\natexlab{c}})}\BibitemShut
  {NoStop}%
\bibitem [{\citenamefont {Liu}\ \emph {et~al.}(1996)\citenamefont {Liu},
  \citenamefont {Erdemir},\ and\ \citenamefont {Meletis}}]{Liu-1996}%
  \BibitemOpen
  \bibfield  {author} {\bibinfo {author} {\bibfnamefont {Y.}~\bibnamefont
  {Liu}}, \bibinfo {author} {\bibfnamefont {A.}~\bibnamefont {Erdemir}}, \ and\
  \bibinfo {author} {\bibfnamefont {E.}~\bibnamefont {Meletis}},\ }\href
  {\doibase https://doi.org/10.1016/S0257-8972(96)03057-5} {\bibfield
  {journal} {\bibinfo  {journal} {Surface and Coatings Technology}\ }\textbf
  {\bibinfo {volume} {86-87}},\ \bibinfo {pages} {564 } (\bibinfo {year}
  {1996})}\BibitemShut {NoStop}%
\bibitem [{\citenamefont {Voevodin}\ \emph {et~al.}(1996)\citenamefont
  {Voevodin}, \citenamefont {Phelps}, \citenamefont {Zabinski},\ and\
  \citenamefont {Donley}}]{Voevodin-1996}%
  \BibitemOpen
  \bibfield  {author} {\bibinfo {author} {\bibfnamefont {A.}~\bibnamefont
  {Voevodin}}, \bibinfo {author} {\bibfnamefont {A.}~\bibnamefont {Phelps}},
  \bibinfo {author} {\bibfnamefont {J.}~\bibnamefont {Zabinski}}, \ and\
  \bibinfo {author} {\bibfnamefont {M.}~\bibnamefont {Donley}},\ }\href
  {\doibase https://doi.org/10.1016/0925-9635(96)00538-9} {\bibfield  {journal}
  {\bibinfo  {journal} {Diamond and Related Materials}\ }\textbf {\bibinfo
  {volume} {5}},\ \bibinfo {pages} {1264 } (\bibinfo {year}
  {1996})}\BibitemShut {NoStop}%
\bibitem [{\citenamefont {Bouchet}\ \emph {et~al.}(2015)\citenamefont
  {Bouchet}, \citenamefont {Matta}, \citenamefont {Vacher}, \citenamefont
  {Le-Mogne}, \citenamefont {Martin}, \citenamefont {von Lautz}, \citenamefont
  {Ma}, \citenamefont {Pastewka}, \citenamefont {Otschik}, \citenamefont
  {Gumbsch},\ and\ \citenamefont {Moseler}}]{DeBarrosBouchet-2015}%
  \BibitemOpen
  \bibfield  {author} {\bibinfo {author} {\bibfnamefont {M.~D.~B.}\
  \bibnamefont {Bouchet}}, \bibinfo {author} {\bibfnamefont {C.}~\bibnamefont
  {Matta}}, \bibinfo {author} {\bibfnamefont {B.}~\bibnamefont {Vacher}},
  \bibinfo {author} {\bibfnamefont {T.}~\bibnamefont {Le-Mogne}}, \bibinfo
  {author} {\bibfnamefont {J.}~\bibnamefont {Martin}}, \bibinfo {author}
  {\bibfnamefont {J.}~\bibnamefont {von Lautz}}, \bibinfo {author}
  {\bibfnamefont {T.}~\bibnamefont {Ma}}, \bibinfo {author} {\bibfnamefont
  {L.}~\bibnamefont {Pastewka}}, \bibinfo {author} {\bibfnamefont
  {J.}~\bibnamefont {Otschik}}, \bibinfo {author} {\bibfnamefont
  {P.}~\bibnamefont {Gumbsch}}, \ and\ \bibinfo {author} {\bibfnamefont
  {M.}~\bibnamefont {Moseler}},\ }\href {\doibase 10.1016/j.carbon.2015.02.041}
  {\bibfield  {journal} {\bibinfo  {journal} {Carbon}\ }\textbf {\bibinfo
  {volume} {87}},\ \bibinfo {pages} {317 } (\bibinfo {year}
  {2015})}\BibitemShut {NoStop}%
\bibitem [{\citenamefont {Pastewka}\ \emph {et~al.}(2010)\citenamefont
  {Pastewka}, \citenamefont {Moser},\ and\ \citenamefont
  {Moseler}}]{Pastewka2010}%
  \BibitemOpen
  \bibfield  {author} {\bibinfo {author} {\bibfnamefont {L.}~\bibnamefont
  {Pastewka}}, \bibinfo {author} {\bibfnamefont {S.}~\bibnamefont {Moser}}, \
  and\ \bibinfo {author} {\bibfnamefont {M.}~\bibnamefont {Moseler}},\ }\href
  {\doibase 10.1007/s11249-009-9566-8} {\bibfield  {journal} {\bibinfo
  {journal} {Tribology Letters}\ }\textbf {\bibinfo {volume} {39}},\ \bibinfo
  {pages} {49} (\bibinfo {year} {2010})}\BibitemShut {NoStop}%
\bibitem [{\citenamefont {Luan}\ and\ \citenamefont
  {Robbins}(2005)}]{Robbins-2005}%
  \BibitemOpen
  \bibfield  {author} {\bibinfo {author} {\bibfnamefont {B.}~\bibnamefont
  {Luan}}\ and\ \bibinfo {author} {\bibfnamefont {M.}~\bibnamefont {Robbins}},\
  }\href {\doibase 10.1038/nature03700} {\bibfield  {journal} {\bibinfo
  {journal} {Nature}\ }\textbf {\bibinfo {volume} {435}},\ \bibinfo {pages}
  {929} (\bibinfo {year} {2005})}\BibitemShut {NoStop}%
\bibitem [{\citenamefont {Ta}\ \emph {et~al.}(2021{\natexlab{b}})\citenamefont
  {Ta}, \citenamefont {Tran}, \citenamefont {Tieu}, \citenamefont {Zhu},
  \citenamefont {Yu},\ and\ \citenamefont {Ta}}]{Huong-2021}%
  \BibitemOpen
  \bibfield  {author} {\bibinfo {author} {\bibfnamefont {H.~T.~T.}\
  \bibnamefont {Ta}}, \bibinfo {author} {\bibfnamefont {N.~V.}\ \bibnamefont
  {Tran}}, \bibinfo {author} {\bibfnamefont {A.~K.}\ \bibnamefont {Tieu}},
  \bibinfo {author} {\bibfnamefont {H.}~\bibnamefont {Zhu}}, \bibinfo {author}
  {\bibfnamefont {H.}~\bibnamefont {Yu}}, \ and\ \bibinfo {author}
  {\bibfnamefont {T.~D.}\ \bibnamefont {Ta}},\ }\href {\doibase
  10.1021/acs.jpcc.1c03725} {\bibfield  {journal} {\bibinfo  {journal} {The
  Journal of Physical Chemistry C}\ }\textbf {\bibinfo {volume} {125}},\
  \bibinfo {pages} {16875} (\bibinfo {year} {2021}{\natexlab{b}})},\ \Eprint
  {http://arxiv.org/abs/https://doi.org/10.1021/acs.jpcc.1c03725}
  {https://doi.org/10.1021/acs.jpcc.1c03725} \BibitemShut {NoStop}%
\bibitem [{\citenamefont {Kokalj}(1999)}]{KOKALJ-1999}%
  \BibitemOpen
  \bibfield  {author} {\bibinfo {author} {\bibfnamefont {A.}~\bibnamefont
  {Kokalj}},\ }\href {\doibase 10.1016/S1093-3263(99)00028-5} {\bibfield
  {journal} {\bibinfo  {journal} {Journal of Molecular Graphics and Modelling}\
  }\textbf {\bibinfo {volume} {17}},\ \bibinfo {pages} {176 } (\bibinfo {year}
  {1999})}\BibitemShut {NoStop}%
\bibitem [{\citenamefont {Kokalj}(2003)}]{KOKALJ-2003}%
  \BibitemOpen
  \bibfield  {author} {\bibinfo {author} {\bibfnamefont {A.}~\bibnamefont
  {Kokalj}},\ }\href {\doibase 10.1016/S0927-0256(03)00104-6} {\bibfield
  {journal} {\bibinfo  {journal} {Computational Materials Science}\ }\textbf
  {\bibinfo {volume} {28}},\ \bibinfo {pages} {155 } (\bibinfo {year}
  {2003})},\ \bibinfo {note} {proceedings of the Symposium on Software
  Development for Process and Materials Design}\BibitemShut {NoStop}%
\bibitem [{\citenamefont {Hunter}(2007)}]{Hunter-2007}%
  \BibitemOpen
  \bibfield  {author} {\bibinfo {author} {\bibfnamefont {J.~D.}\ \bibnamefont
  {Hunter}},\ }\href {\doibase 10.1109/MCSE.2007.55} {\bibfield  {journal}
  {\bibinfo  {journal} {Computing In Science \& Engineering}\ }\textbf
  {\bibinfo {volume} {9}},\ \bibinfo {pages} {90} (\bibinfo {year}
  {2007})}\BibitemShut {NoStop}%
\end{thebibliography}%

%\end{thebibliography}

\end{document}